\documentclass[twocolumn,showpacs,prb,aps]{revtex4}
\usepackage{amsmath}
\usepackage{graphicx}
\usepackage{dcolumn}
\usepackage{color}
\newcommand{\be}{\begin{equation}}
\newcommand{\ee}{\end{equation}}
\newcommand{\bea}{\begin{eqnarray}}
\newcommand{\eea}{\end{eqnarray}}
\newcommand{\bs}{\begin{split}}
\newcommand{\bse}{\begin{subequations}}
\newcommand{\ese}{\end{subequations}}

\begin{document}
\title{Superconducting and Normal State Properties of $A$Pd$_2$As$_2$ ($A$ = Ca, Sr, Ba) Single Crystals}
\author{V. K. Anand}
\altaffiliation{vanand@ameslab.gov}
\author{H. Kim}
\author{M. A. Tanatar}
\author{R. Prozorov}
\author{D. C. Johnston}
\altaffiliation{johnston@ameslab.gov}
\affiliation {Ames Laboratory and Department of Physics and Astronomy, Iowa State University, Ames, Iowa 50011}

\date{May 29, 2013}

\begin{abstract}

The synthesis and crystallography, magnetic susceptibility $\chi$, magnetization $M$, specific heat $C_{\rm p}$, in-plane electrical resistivity $\rho$ and in-plane magnetic penetration depth measurements are reported for single crystals of $A$Pd$_2$As$_2$ ($A$ = Ca, Sr, Ba) versus temperature $T$ and magnetic field $H$\@.  The crystals were grown using PdAs self-flux.  CaPd$_2$As$_2$ and SrPd$_2$As$_2$ crystallize in a collapsed body-centered tetragonal ${\rm ThCr_2Si_2}$-type structure ($I4/mmm$), whereas BaPd$_2$As$_2$ crystallizes in the primitive tetragonal CeMg$_2$Si$_2$-type structure ($P4/mmm$), in agreement with literature data.  The $\rho (T)$ data exhibit metallic behavior for all three compounds.  Bulk superconductivity is reported for CaPd$_2$As$_2$ and SrPd$_2$As$_2$ below $T_{\rm c} = 1.27$ and 0.92~K, respectively, whereas only a trace of superconductivity is found in BaPd$_2$As$_2$.  No other phase transitions were observed. The $\chi(T)$ and $M(H)$ data reveal anisotropic diamagnetism in the normal state, with $\chi_{c} > \chi_{ab}$ for CaPd$_2$As$_2$ and BaPd$_2$As$_2$, and $\chi_{c} < \chi_{ab}$ for SrPd$_2$As$_2$. The normal and superconducting state data indicate that CaPd$_2$As$_2$ and SrPd$_2$As$_2$ are conventional type-II nodeless $s$-wave electron-phonon superconductors.  The electronic superconducting-state heat capacity data for CaPd$_2$As$_2$, which has an extremely sharp heat capacity jump at $T_{\rm c}$, are analyzed using our recent elaboration of the $\alpha$-model of the BCS theory of superconductivity, which indicates that the $s$-wave gap in this compound is anisotropic in momentum space.

\end{abstract}


\maketitle

\section{\label{Intro} Introduction}

The observation of high-temperature superconductivity (SC) with transition temperatures $T_{\rm c} \lesssim  38$~K in doped 122-type iron arsenides such as in $A_{1-x}$K$_x$Fe$_2$As$_2$ ($A$ = Ba, Sr, Ca and Eu) compounds has stimulated great interest in these materials. \cite{Rotter2008a, Chen2008a, Sasmal2008, Wu2008, Jeevan2008, Sefat2008, Torikachvili2008, Ishida2009, Alireza2009, Wang2009, Johnston2010, Canfield2010, Mandrus2010, Paglione2010, Dai2012} The parent compounds $A$Fe$_2$As$_2$ with $A$ = Ca, Sr, and Ba are itinerant antiferromagnetic (AF) semimetals that undergo a structural distortion from a tetragonal structure to an orthorhombic one on cooling below room temperature.  The structural transition precedes a long-range AF itinerant spin density wave (SDW) transition. Superconductivity in these compounds emerges upon suppression of the SDW transition that can be achieved either by partial chemical substitutions at either the $A$, Fe or As sites or by application of external pressure.\cite{Johnston2010} The same phenomenology is found in the high-$T_{\rm c}$ cuprates where the long-range AF order must be largely suppressed by doping prior to the emergence of superconductivity, but strong dynamic AF spin fluctuations must still be present.  Thus, the iron arsenides and the high-$T_{\rm c}$ cuprates have the same generic phase diagram for the emergence of superconductivity, even though the cuprate parent compounds are AF insulators rather than SDW semimetals.\cite{Johnston2010, Canfield2010, Mandrus2010, Johnston1997, Damascelli2003, Lee2006}

The partial substitutions at the Fe-site in $A$Fe$_2$As$_2$ by transition metals $M$ in Ba(Fe$_{1-x}M_x$)$_2$As$_2$ with $M$ = Cr, Mn, Co, Ni, Cu, Ru, Rh, and Pd have recently been studied both theoretically and experimentally from the perspective of the degree of charge doping because of the changes in the magnetic and SC properties caused by such substitutions. \cite{Canfield2010, Canfield2009, Kasinathan2009, Sefat2008, WangC2009, Li2009, Sefat2009, Marty2011, Liu2010, Kim2010, Ni2009, Han2009, Thaler2010, Dhaka2011, Wadati2010, Berlijn2012, LiuC2010, LiuC2011, Neupane2011, Bittar2011, McLeod2012, Levy2012, Ideta2012}  The substitutions for the Fe atoms by the $3d$ elements Co and Ni and the $4d$ elements Rh and Pd which have higher number of outer-shell $d$-electrons than that of Fe are found to induce superconductivity. \cite{Sefat2008, WangC2009, Li2009,Ni2009, Han2009}  However, no superconductivity is induced by Mn or Cr substitutions having a lower number of 3$d$ electrons. \cite{Sefat2009, Marty2011, Liu2010, Kim2010} On the other hand isoelectronic substitution of Fe by the 4$d$ transition metal Ru is also found to induce the superconductivity. \cite{Thaler2010, Dhaka2011} The common feature in the SC materials is that the long-range SDW order in the parent compounds must be largely suppressed before SC occurs as noted above, where, in addition, the AF spin fluctuations are still strong.  The latter fact suggests an unconventional magnetic mechanism for the SC in the FeAs-based materials.  The extent and effects of electron and hole doping by such substitutions are still being debated.\cite{Canfield2010, Wadati2010, McLeod2012, Ideta2012, Anand2012a, Anand2012b}

While the effect of partial substitutions of Fe by transition metals have been investigated extensively, the physical properties of the end-point compounds $AM_2$As$_2$ of the $A$(Fe$_{1-x}M_x$)$_2$As$_2$ series are often not known in detail. Recently we reported crystallography and physical property studies of the end-point compounds ${\rm SrCu_2As_2}$ and ${\rm CaCu_{1.7}As_2}$ for $M$ = Cu which are $sp$-band metals.\cite{Anand2012a,Anand2012b} These compounds form in a collapsed tetragonal (cT) structure in which the formal oxidation state of As is As$^{-2} \equiv $ [As--As]$^{-4}$/2 due to the interlayer As--As bonding in the cT phase. Thus with the formal oxidation states of Sr and Ca taken to be $+2$, the Cu atoms in these cT compounds are then  Cu$^{+1}$ with a  nonmagnetic $3d^{10}$ configuration as observed\cite{Anand2012a, Anand2012b} and also as previously predicted from electronic structure calculations for ${\rm SrCu_2As_2}$ by Singh.\cite{Singh2009}  Thus, extrapolating from high to low Cu concentrations, the electron or hole doping effect resulting from substitution of $M$ = Cu for Fe in $A$(Fe$_{1-x}M_x$)$_2$As$_2$ compounds is ambiguous and interesting.\cite{Canfield2010,Ideta2012,Anand2012a}

\begin{figure}
\includegraphics[width=3.3in]{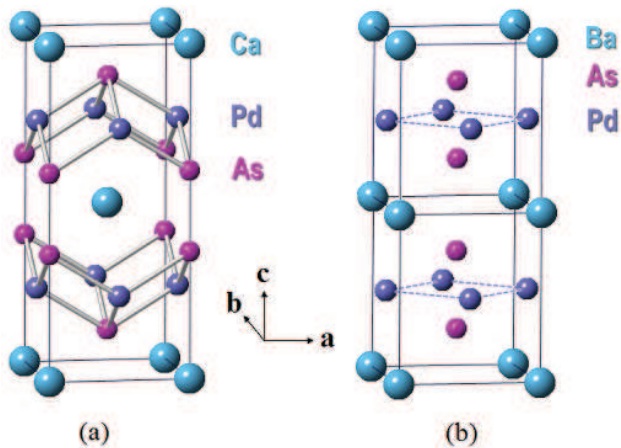}
\caption{(Color online) (a) A unit cell of ThCr$_2$Si$_2$-type body-centered tetragonal crystal structure ($I4/mmm$) of CaPd$_2$As$_2$ and SrPd$_2$As$_2$. (b) Two unit cells, stacked along the $c$-axis, of the CeMg$_2$Si$_2$-type primitive tetragonal crystal structure ($P4/mmm$) of BaPd$_2$As$_2$.}
\label{fig:APd2As2_structure}
\end{figure}

To provide insight into the $T$-$x$ phase diagrams of $A$(Fe$_{1-x}M_x$)$_2$As$_2$ systems with $M$ = Pd, we  report here our investigations of the crystallography and physical properties of single crystals of the three (Ca,Sr,Ba)Pd$_2$As$_2$ end-point compounds.  ${\rm CaPd_2As_2}$ and ${\rm SrPd_2As_2}$ are reported to form in the body-centered tetragonal ${\rm ThCr_2Si_2}$-type structure (space group $I4/mmm$) and ${\rm BaPd_2As_2}$ in the primitive tetragonal ${\rm CeMg_2Si_2}$-type structure (space group $P4/mmm$),\cite{Mewis1984, Hofman1985} as shown in Fig.~\ref{fig:APd2As2_structure} for ${\rm CaPd_2As_2}$ and ${\rm BaPd_2As_2}$.  Both structures contain similar Pd square-lattices. However, while the Ca atoms in ${\rm CaPd_2As_2}$ form a body-centered tetragonal sublattice, in the ${\rm BaPd_2As_2}$ structure the Ba atoms form a primitive tetragonal sublattice.  Furthermore, the fourfold coordination of Pd by As is tetrahedral in ${\rm CaPd_2As_2}$ but is planar rectangular in ${\rm BaPd_2As_2}$.  The ${\rm BaPd_2As_2}$ structure can be derived from that of ${\rm CaPd_2As_2}$ by a \mbox{[$\frac{1}{2}$, $\frac{1}{2}$, 0]} translation of the central As-Ca-As layer.  Superconductivity was recently reported below $T_{\rm c} = 3.0$~K in the similar compound ${\rm SrPd_2Ge_2}$ with the ThCr$_2$Si$_2$-type structure.\cite{Fujii2009}

In this paper we report crystallographic investigations of crushed single crystal powders and physical property measurements of single crystals of the three (Ca,Sr,Ba)Pd$_2$As$_2$ compounds using  magnetic susceptibility $\chi$, isothermal magnetization $M$, specific heat $C_{\rm p}$, and $ab$-plane electrical resistivity $\rho$ measurements as a function of temperature $T$ and applied magnetic field $H$\@.  The published structures of these compounds are confirmed.  The $\rho(T)$, $\chi(T)$ and $C_{\rm p}(T)$ data demonstrate that all three compounds are $sp$-band-like metals. The $\chi (T)$ data indicate anisotropic diamagnetic behavior in all three compounds with $\chi_{c} > \chi_{ab}$ for CaPd$_2$As$_2$ and BaPd$_2$As$_2$ and $\chi_{c} < \chi_{ab}$ for SrPd$_2$As$_2$.  Bulk superconducting transitions at $T_{\rm c} = 1.27(3)$~K for ${\rm CaPd_2As_2}$ and $T_{\rm c} = 0.92(5)$~K for SrPd$_2$As$_2$ are also reported together with other superconducting-state properties derived from the above measurements as well as from magnetic penetration depth measurements.  We analyzed the superconducting-state electronic entropy and heat capacity data using our recent elaboration of the $\alpha$-model of the BCS theory of superconductivity.\cite{Johnston2013} The data indicate that ${\rm CaPd_2As_2}$ and SrPd$_2$As$_2$ are both conventional nodeless type-II electron-phonon superconductors, but with anisotropic $s$-wave gaps.  Filamentary superconductivity was also detected below 2.0~K in BaPd$_2$As$_2$.

The experimental details are described in Sec.~\ref{ExpDetails}.  The crystallographic studies of CaPd$_2$As$_2$, SrPd$_2$As$_2$ and BaPd$_2$As$_2$are presented in Sec.~\ref{Crystallography}.  The superconducting and normal state properties of CaPd$_2$As$_2$ and SrPd$_2$As$_2$ are given in Secs.~\ref{CaPd2As2} and \ref{SrPd2As2}, respectively, and the measurements of their superconducting magnetic penetration depths versus temperature are presented in Sec.~\ref{Sec:LondonPD}.  The normal state properties of BaPd$_2$As$_2$ are presented in Sec.~\ref{BaPd2As2}.  A summary of the experimental results and analyses and the conclusions are given in Sec.~\ref{Conclusion}.

\section{\label{ExpDetails} EXPERIMENTAL DETAILS}

Single crystals of $A$Pd$_2$As$_2$ ($A$ = Ba, Ca, Sr) were grown using PdAs flux. The starting materials were high-purity elemental Ca (99.98\%), Pd (99.998\%) and As (99.99999\%) from Alfa Aesar and Sr (99.95\%) and Ba (99.99\%) from Sigma Aldrich. The Ca, Sr or Ba and prereacted PdAs flux taken in a 1:4 molar ratio were placed in alumina crucibles and sealed inside evacuated fused silica tubes. The sealed samples were heated to 1100~$^\circ$C at a rate of 60--80~$^\circ$C/h and held there for 12~h. After cooling at a rate of 2.5~$^\circ$C/h to 800~$^\circ$C, shiny plate-like crystals of typical size $2 \times 1.5 \times 0.4$~mm$^3$ were separated by decanting the flux with a centrifuge at that temperature.

The phase purity and the chemical composition of the CaPd$_2$As$_2$, SrPd$_2$As$_2$ and BaPd$_2$As$_2$ crystals were measured using a JEOL scanning electron microscope (SEM) equipped with an energy-dispersive x-ray analyzer (EDX). High-resolution SEM images demonstrated that the crystals were single-phase.  The EDX analyses of two or three single crystals of each compound were consistent with the ideal 1~:~2~:~2 stoichiometry.  The crystal structures of the samples were determined by Rietveld refinements of powder x-ray diffraction (XRD) data for crushed single crystals collected on a Rigaku Geigerflex x-ray diffractometer using Cu K$_\alpha$ radiation.  The \mbox{Rietveld} refinements were carried out using the {\tt FullProf} software\cite{Rodriguez1993} which also confirmed the single-phase nature of the crystals.

The magnetization measurements were performed using a Quantum Design, Inc., superconducting quantum interference device magnetic properties measurement system (MPMS).  We use Gaussian cgs units for the magnetic results and discussion, where the unit of magnetic field $H$ is the Oe = G, but we also use the Tesla, with 1~T $\equiv  10^4$~Oe, as a unit of convenience (the magnetic moment output by the MPMS software is in Gaussian cgs units where ${\rm 1~emu = 1~G\,cm^3}$).\cite{Johnston2010}  In order to obtain the magnetic moment of the crystals in a particular field $H$ we subtracted the pre-calibrated magnetic moment of the sample holder (quartz rod/plastic piece and GE varnish that were used to mount the sample for MPMS measurements) in the same $H$\@.  Due to the small magnitudes of the magnetic moments of the small crystals and the uncertainty in the magnetic moment of sample holder, the accuracy of the $M(H,T)$ and $\chi(T)$ data of the samples reported here is of order 10\%.

The heat capacity measurements were carried out using the relaxation method in a Quantum Design, Inc., physical properties measurement system (PPMS)\@. Temperatures down to 0.45~K were obtained using a $^3$He attachment to the PPMS\@.  The $ab$-plane $\rho(T)$ measurements were performed by the standard four-probe ac technique using the ac transport option of the PPMS\@.  The electrical leads were 25~$\mu$m diameter platinum wires attached to the crystals with EPO-TEK P1011 silver epoxy cured at 110 $^\circ$C for one hour.  The accuracy of $\rho$ is of order 10\% due to uncertainties in the geometric factor.

The temperature variation of the magnetic penetration depth $\lambda_{\rm eff}$ was measured using a tunnel diode resonator (TDR) technique operating at about 15 MHz. The resonator consists of an $LC$ tank circuit with a single-layer coil of inductance $L \sim 1~\mu$H, a capacitor with capacitance $C \sim 100$~pF and a tunnel diode that is biased to the region of negative differential resistance, thus compensating the losses in the circuit. The circuit, therefore, self-oscillates at a frequency $f=(2 \pi \sqrt{LC})^{-1}$. When a sample with magnetic susceptibility $\chi$ is inserted into the coil, the total inductance changes and the resonant frequency shifts accordingly by an amount which is proportional to $\chi$. The temperature dependence of the resonance frequency shift $\Delta f(T)$ induced by changes in the sample's magnetic response is related to the magnetic susceptibility $\chi$ and hence $\lambda_{\rm eff}(T)$ by \cite{Prozorov2000,Prozorov2006,Prozorov2000a}
\bea
\Delta f(T)&=&-4\pi\chi(T)G\\*
&\approx& -G\left\{1-\frac{\lambda_{\rm eff}(T)}{L_{\rm eff}}\tanh\left[\frac{L_{\rm eff}}{\lambda_{\rm eff}(T)}\right]\right\}\nonumber,
\eea
where $G$ is a sample- and coil-dependent calibration constant and $L_{\rm eff}$ is the effective sample dimension.~\cite{Prozorov2000} The value of $G$ is determined experimentally by pulling the sample out of the coil \emph{in situ} and measuring the total associated frequency shift, $\Delta f_0$, so that $\Delta\lambda_{\rm eff}(T)=L_{\rm eff}\ \delta f(T)/\Delta f_0$ where $\Delta\lambda_{\rm eff}(T)=\lambda_{\rm eff}(T)-\lambda_{\rm eff}(T_{\rm min})$, $\delta f(T)=f(T)-f(T_{\rm min})$ and $T_{\rm min}$ is the minimum temperature of the measurement.\cite{Prozorov2000}  Another way to determine $G$ is by matching the temperature dependence of the skin depth, $d(T)$, obtained from the resonator response in the normal state to the measured resistivity, $\rho(T)$, by using the relation $d(T)=(c/2\pi)\sqrt{\rho(T)/f}$. \cite{Hardy1993}  The ac magnetic field was applied along the $c$-axis of single crystals of CaPd$_2$As$_2$ and SrPd$_2$As$_2$, so the reported penetration depths are the respective values in the $ab$-plane.

\section{\label {Crystallography} Crystallography}
\begin{figure}
\includegraphics[width=3in]{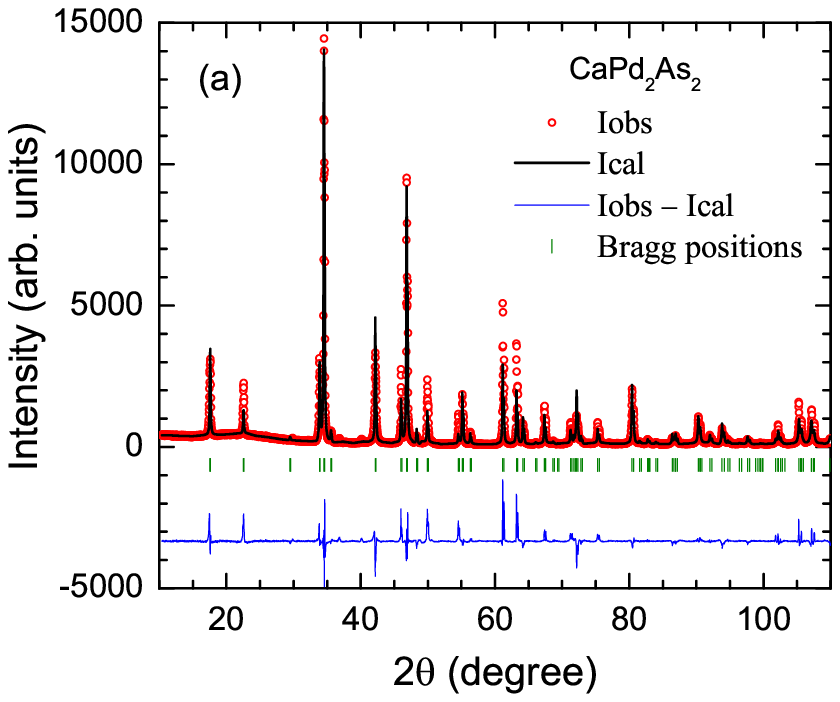}
\includegraphics[width=3in]{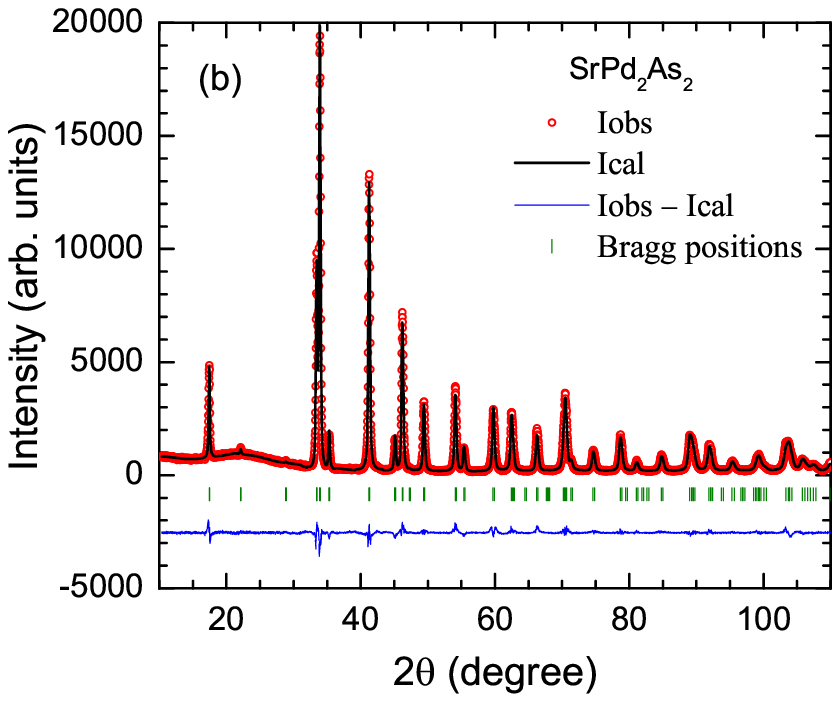}
\includegraphics[width=3in]{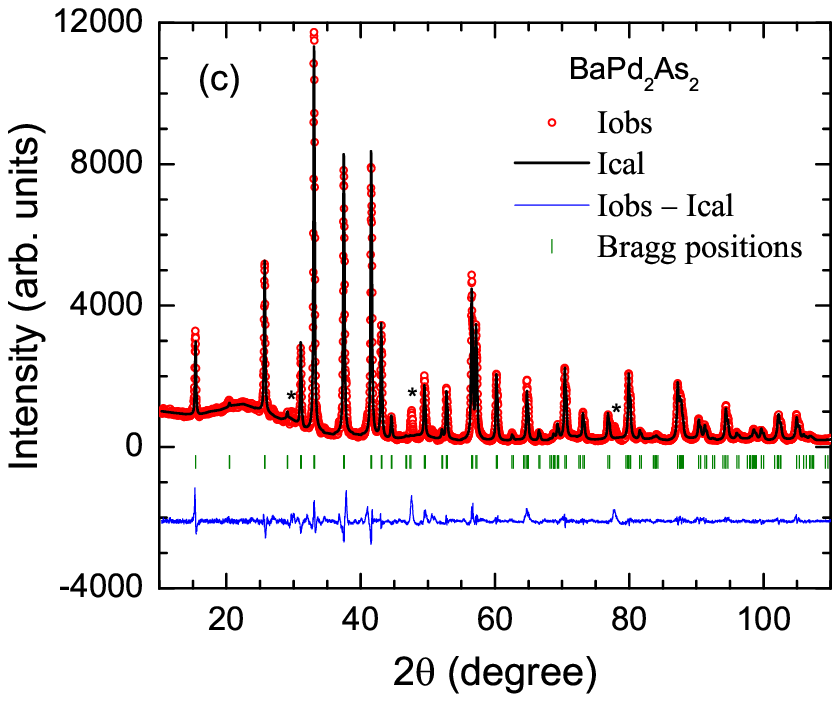}
\caption{\label{fig:APd2As2_XRD} (Color online) Powder x-ray diffraction patterns of (a) CaPd$_2$As$_2$, (b) SrPd$_2$As$_2$ and (c) BaPd$_2$As$_2$ recorded at room temperature. The solid lines through the experimental points are the Rietveld refinement profiles calculated [(a) and (b)] for the ThCr$_2$Si$_2$-type body-centered tetragonal structure (space group $I4/mmm$), and (c) for the CeMg$_2$Si$_2$-type primitive tetragonal structure (space group $P4/mmm$).  In (a), (b) and (c), the short vertical bars mark the fitted Bragg peak positions. The lowermost curves represent the differences between the experimental and calculated intensities. The unindexed peaks marked with stars correspond to peaks from residual PdAs flux on the surface of the samples.}
\end{figure}

The room temperature powder XRD patterns for crushed single crystals of CaPd$_2$As$_2$, SrPd$_2$As$_2$ and BaPd$_2$As$_2$ are shown in Figs.~\ref{fig:APd2As2_XRD}(a), \ref{fig:APd2As2_XRD}(b) and \ref{fig:APd2As2_XRD}(c), respectively.  The weak unindexed peaks marked with stars for BaPd$_2$As$_2$ arise from the small amount of flux attached to the surface of the crystals prior to crushing them for the measurements.  The Rietveld refinements confirmed the reported ${\rm ThCr_2Si_2}$-type body-centered tetragonal structure (space group $I4/mmm$) of ${\rm CaPd_2As_2}$ and SrPd$_2$As$_2$ and the CeMg$_2$Si$_2$-type primitive tetragonal structure (space group $P4/mmm$) of BaPd$_2$As$_2$.\cite{Mewis1984,Hofman1985}

The refinement profiles for these structural models are shown in Fig.~\ref{fig:APd2As2_XRD}. During the final refinements, the thermal parameters $B$ were set to $B = 0$ and the occupancies of the respective atomic sites were fixed to unity, since there were no improvements in the quality of fits or significant changes in the lattice parameters or in the As $c$-axis position parameter $z_{\rm As}$ upon making small changes in $B$ and in the occupancies.  The crystallographic and refinement parameters are listed in Tables~\ref{tab:XRD1} and \ref{tab:XRD2}.  For comparison the lattice parameters and $z_{\rm As}$ values from the literature are also listed in Table~\ref{tab:XRD1}, where good agreement is found.

The $c/a$ ratio and the interlayer As--As distance $d_{\rm As-As} = (1-2z_{\rm As})c$ for (Ca,Sr)Pd$_2$As$_2$ and $2z_{\rm As}c$ for BaPd$_2$As$_2$ are listed in Table~\ref{tab:XRD1}, where a ``layer'' is defined as an As--Pd--As slab in Fig.~\ref{fig:APd2As2_structure}. The values of $d_{\rm As-As}$ for CaPd$_2$As$_2$ and SrPd$_2$As$_2$ are close to the covalent single-bond distance of 2.38~\AA\ for As,\cite{Cordero2008} indicating that these two compounds have cT structures.\cite{Anand2012a}  Therefore, as discussed in Sec.~\ref{Intro}, the formal oxidation state of Pd in these (Ca,Sr)Pd$_2$As$_2$ compounds is Pd$^{+1}$, which is the same formal oxidation state as for Cu in ${\rm ThCr_2Si_2}$-type ${\rm SrCu_2As_2}$.\cite{Anand2012a}

\begin{table*}
\caption{\label{tab:XRD1}  Crystallographic and Rietveld refinement parameters obtained from powder XRD data for crushed $A$Pd$_2$As$_2$ ($A$ = Ba, Ca, Sr) crystals. Error bars for the last digit of a quantity are given in parentheses and literature references are given in square brackets.}
\begin{ruledtabular}
\begin{tabular}{llll}
 & CaPd$_2$As$_2$ & SrPd$_2$As$_2$ & BaPd$_2$As$_2$ \\
\hline
Structure & ThCr$_2$Si$_2$-type & ThCr$_2$Si$_2$-type & CeMg$_2$Si$_2$-type  \\
 & body-centered tetragonal & body-centered  tetragonal & primitive tetragonal \\
Space group & $I4/mmm$ & $I4/mmm$ & $P4/mmm$ \\
Formula units/unit cell ($Z$) & 2 & 2 & 1 \\
Lattice parameters\\
  \hspace{0.8 cm}  $a$ (\AA)            			&   4.2824(2)  & 4.3759(1)  & 4.3438(2) \\	
													&   4.299(1) [\onlinecite{Mewis1984}]  & 4.383(1) [\onlinecite{Mewis1984}] & 4.346(1) [\onlinecite{Mewis1984}] \\
													&   4.283(1) [\onlinecite{Hofman1985}]  & 4.380(2) [\onlinecite{Hofman1985}] &  \\
  \hspace{0.8 cm}  $c$ (\AA)            			&   10.0880(4) & 10.1671(3) & 5.7536(2) \\	
													&   10.102(2) [\onlinecite{Mewis1984}] & 10.179(2) [\onlinecite{Mewis1984}] & 5.758(1) [\onlinecite{Mewis1984}] \\
													&   10.093(1) [\onlinecite{Hofman1985}] & 10.169(1) [\onlinecite{Hofman1985}] &  \\
    \hspace{0.8 cm} $c/a$						&   2.3557(6)	   &  2.3234(4)      & 1.3246(4) \\
  \hspace{0.8 cm}  $V_{\rm cell}$  ({\AA}$^{3}$) 	&  185.01(1)   & 194.69(1)  & 108.56(1)  \\
As $c$-axis coordinate $z_{\rm As}$ 						& 0.3763(3)    & 0.3768(1)  & 0.2705(4) \\
													& 0.3796(2) [\onlinecite{Mewis1984}]    & 0.3766(2) [\onlinecite{Mewis1984}]  & 0.2700(8) [\onlinecite{Mewis1984}] \\
													& 			   & 0.3768(1) [\onlinecite{Hofman1985}] &  \\
As--As interlayer bond  			& 2.496(7)  	   & 2.505(2)  	& 3.113(4) \\
\hspace{0.35in}distance $d_{\rm As-As}$ (\AA)\\
Theoretical density $({\rm g/cm^3})$			&		7.230		&	7.682	& 	7.648\\
Molar volume $V_{\rm M}~({\rm cm^3/mol}$)					&	55.71		&	58.62	&	65.38	\\
Refinement quality \\
  \hspace{0.8 cm}    $\chi^2$ (\%)	   & 11.8 & 3.68 & 8.09 \\	
  \hspace{0.8 cm}    $R_{\rm p}$ (\%)  & 12.7 & 5.41 & 7.39 \\
  \hspace{0.8 cm}    $R_{\rm wp}$ (\%) & 18.6 & 7.23 & 11.2 \\
\end{tabular}
\end{ruledtabular}
\end{table*}

\begin{table}
\caption{\label{tab:XRD2} Atomic coordinates obtained from the Rietveld refinements of powder XRD data for crushed $A$Pd$_2$As$_2$ ($A$ = Ba, Ca, Sr) crystals.}
\begin{ruledtabular}
\begin{tabular}{lcccc}
   \hspace{0.6cm}  Atom & Wyckoff   &	 $x$ 	&	$y$	&	$z$	  \\	
   & symbol & \\
\hline

CaPd$_2$As$_2$ ($I4/mmm$)\\				
   \hspace{0.8cm}     Ca  & 2a  	&	 0 	&	0	&   0		  \\
   \hspace{0.8cm}     Pd  & 4d	&	 0 	&	1/2	&	1/4 	  		\\
   \hspace{0.8cm}     As  & 4e 	&	 0 	&   0 	&	0.3763(3)         \\

SrPd$_2$As$_2$ ($I4/mmm$)\\				
   \hspace{0.8cm}     Sr  & 2a  	&	 0 	&	0	&   0		  \\
   \hspace{0.8cm}     Pd  & 4d	&	 0 	&	1/2	&	1/4 	  		\\
   \hspace{0.8cm}     As  & 4e 	&	 0 	&   0 	&	0.3768(1)         \\

BaPd$_2$As$_2$ ($P4/mmm$)\\				
   \hspace{0.8cm}     Ba  & 1a  	&	 0 	&	0	&   0		 	 \\
   \hspace{0.8cm}     Pd  & 2e		&	 0 	&	1/2	&	1/2	  			\\
   \hspace{0.8cm}     As  & 2h 		& 1/2 	&   1/2	&	0.2705(4)    		\\

\end{tabular}
\end{ruledtabular}
\end{table}

\section{\label{CaPd2As2} Physical Properties of C\lowercase{a}P\lowercase{d}$_2$A\lowercase{s}$_2$ Crystals}

\subsection{\label{Sec:CaPd2As2_Rho} Electrical Resistivity}

\begin{figure}
\includegraphics[width=3.3in]{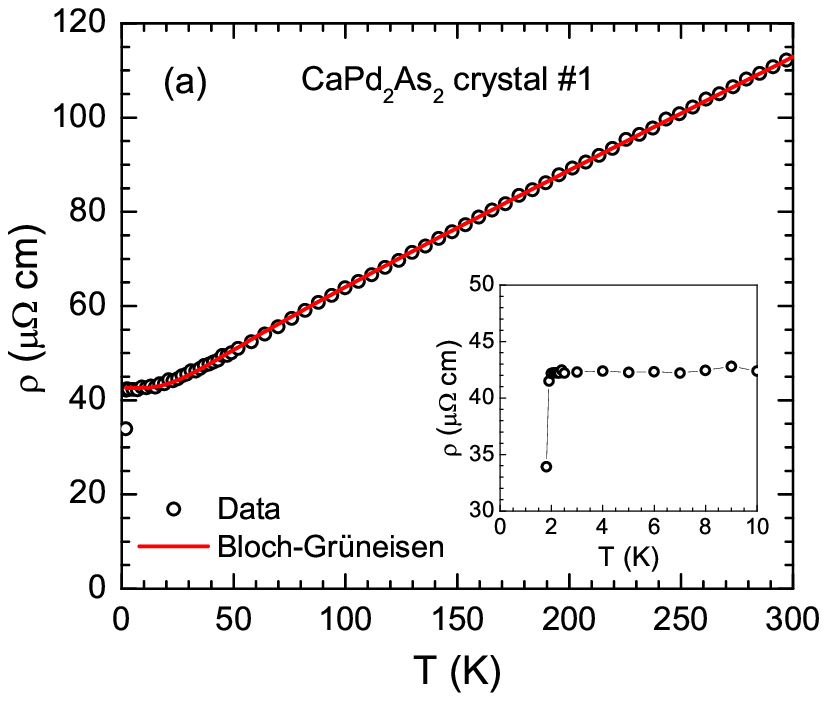}
\includegraphics[width=3.3in]{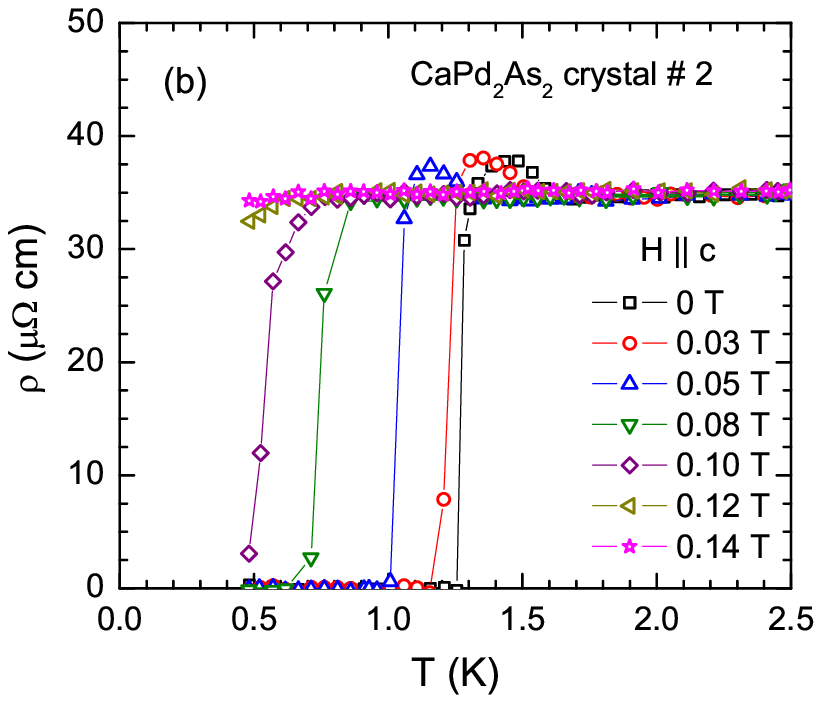}
\caption{(Color online) (a) In-plane electrical resistivity $\rho$ of a CaPd$_2$As$_2$ crystal~\#1 versus temperature~$T$ measured in applied magnetic field $H=0$. The red curve is a fit by the Bloch-Gr\"{u}neisen model in the $T$~range  2--300~K\@. Inset: Expanded plot of $\rho(T)$ below 10~K to show the onset of superconductivity at 2.0~K\@. (b) Expanded plot of $\rho(T)$ of CaPd$_2$As$_2$ crystal~\#2 for 0.45~K~$\leq T \leq$~2.5~K showing the superconducting transition for different values of $H$, as listed, applied along the $c$-axis.}
\label{fig:CaPd2As2_Rho}
\end{figure}

The in-plane $\rho$ of CaPd$_2$As$_2$ versus $T$ is shown for crystal~\#1 at $H=0$ in Fig.~\ref{fig:CaPd2As2_Rho}(a) and for crystal~\#2 under various $H$ in Fig.~\ref{fig:CaPd2As2_Rho}(b). A sharp zero-field superconducting transition is observed in Fig.~\ref{fig:CaPd2As2_Rho}(b) at $T_{\rm c} = 1.28(3)$~K that is suppressed by $H = 0.14$~T to below 0.45~K\@. For crystal~\#1  the superconducting onset is also at 1.8~K as shown in the inset of Fig.~\ref{fig:CaPd2As2_Rho}(a). A broad superconducting transition was seen in  another crystal (\#3) with an onset at 1.8~K and zero resistance at 1.3~K (not shown).  Interestingly the $\rho(T)$ of crystal~\#2 in Fig.~\ref{fig:CaPd2As2_Rho}(b) exhibits an upturn and peak before the resistivity drops due to superconductivity which is suppressed by a field $H \geq 0.08$~T\@. The origin of this $\rho(T)$ peak is not clear.  For crystal~\#1 the residual resistivity at 2~K before entering the superconducting state is $\rho_0 = 43~\mu \Omega$\,cm and the residual resistivity ratio is \mbox{RRR~$\equiv \rho(300\,{\rm K}) / \rho(2\,{\rm K}) \approx 3$}.

In the following we fit the normal-state $\rho(T)$ data by the Bloch-Gr\"{u}neisen (BG) model which describes the electrical resistivity $\rho_{\rm BG}(T)$ due to scattering of the charge carriers by longitudinal acoustic lattice vibrations in the absence of Umklapp scattering, given by \cite{Blatt1968, Ryan2012}
\bse
\label{Eqs:BGModel}
\begin{equation}
\rho_{\rm {BG}}(T/\Theta _{\rm R})= 4 \mathcal{R} \left( \frac{T}{\Theta _{\rm R}}\right)^5 \int_0^{\Theta_{\rm{R}}/T}{\frac{x^5}{(e^x-1)(1-e^{-x})}dx},
\label{eq:Bloch-Gruneisen}
\end{equation}
where $\Theta_{\rm R}$ is the Debye temperature obtained from fitting resistivity measurements, $\mathcal{R}$ is a material-dependent prefactor that is independent of $T$ and 
\be
\rho_{\rm B}(T/\Theta_R = 1) = 0.946\,463\,5\,{\cal R}.
\label{Eq:t=1}
\ee
The experimental $\rho(T)$ data are fitted by
\begin{equation}
\rho(T) = \rho_0 + \rho(\Theta_{\rm R}) \rho_{\rm n}(T/\Theta_{\rm R}),
\label{eq:BG_fit}
\end{equation}
where Eq.~(\ref{eq:Bloch-Gruneisen}) yields the normalized dimensionless BG resistivity
\bea
\rho_{\rm n}(T/\Theta_{\rm R}) &=& 4.226\,259 \left( \frac{T}{\Theta _{\rm R}}\right)^5 \label{eq:BG_R}\\*
&& \times\int_0^{\Theta_{\rm{R}}/T} {\frac{x^5}{(e^x-1)(1-e^{-x})}dx}. \nonumber
\eea
\ese

\begin{table}
\caption{\label{Tab:RhoFitParams} Parameters derived from Bloch-Gr\"uneisen fits to the resistivities $\rho$ within the $ab$-plane of the listed single crystals obtained using Eqs.~(\ref{Eqs:BGModel}).  Here $\rho_0$ is the residual resistivity extrapolated to $T=0$, $\Theta_{\rm R}$ is the Debye temperature determined from resistivity measurements, $\rho(\Theta_{\rm R})$ is the fitted value of $\rho$ at $T = \Theta_{\rm R}$, and ${\cal R}$ is obtained from the fitted value of $\rho(\Theta_{\rm R})$ using Eq.~(\ref{Eq:t=1}).  The $\rho$ and ${\cal R}$ values do not take into account the systematic error of order 10\% arising from uncertainties in the geometric factor required to calculate $\rho$ of a crystal from its measured resistance.  The accuracy of $\Theta_{\rm R}$ is not affected by this systematic error since this parameter is determined solely from the $T$~dependence of~$\rho$.}
\begin{ruledtabular}
\begin{tabular}{lcccc}
Compound 		& $\rho_0$ 		&  $\Theta_{\rm R}$ 	&	$\rho(\Theta_{\rm R})$  	& ${\cal R}$\\
			& ($\mu\Omega$\,cm) & (K)				& ($\mu\Omega$\,cm)			& ($\mu\Omega$\,cm)\\
\hline
CaPd$_2$As$_2$ &  42.60(7)  	&  135(3)  			& 30.3(5) 				&  32.0 \\				
SrPd$_2$As$_2$ &  7.57(6)		&  170(3)  			& 27.6(5)				&  29.2 \\
BaPd$_2$As$_2$ &  1.02(1)		&  114(1)  			& 8.84(1)				&  9.34  \\
\end{tabular}
\end{ruledtabular}
\end{table}

An excellent fit of the $\rho(T)$ data by Eqs.~(\ref{Eqs:BGModel}) was obtained using the three independent fitting parameters $\rho_0$, $\rho(\Theta_{\rm R})$ and $\Theta_{\rm R}$ for 2~K~$\leq T \leq$~300~K, as shown by the red curve in Fig.~\ref{fig:CaPd2As2_Rho}(a). The single parameter $\Theta_{\rm R}$ determines the $T$~dependence of the fit.  While fitting the $\rho(T)$ data we used the analytic Pad\'e approximant function of $T/\Theta _{\rm R}$ in place of Eq.~(\ref{eq:BG_R}) as developed in Ref.~\onlinecite{Ryan2012} which accurately describes $\rho_{\rm n}(T/\Theta _{\rm R})$ and greatly simplifies least-squares fitting of experimental $\rho(T)$ data by the BG theory.  The fitting parameters are summarized in Table~\ref{Tab:RhoFitParams} along with those of the other two compounds discussed below.  The $\mathcal{R}$ value is obtained from the fitted value of $\rho(\Theta_{\rm{R}})$ using Eq.~(\ref{Eq:t=1}).

\subsection{\label{Sec:CaPd2As2_HC} Heat Capacity}

\subsubsection{Overview of the Superconducting and Normal State Heat Capacity}

\begin{table*}
\caption{\label{tab:HCFitParams} The linear specific heat coefficients $\gamma_{\rm n}$ and the coefficients $\beta $ and~$\delta$ of the $T^3$ and $T^5$ terms in the low-$T$ heat capacity, respectively, and the density of states at the Fermi energy ${\cal D}_C(E_{\rm F})$ and ${\cal D}_{\rm band}(E_{\rm F})$ for both spin directions for CaPd$_2$As$_2$, SrPd$_2$As$_2$ and BaPd$_2$As$_2$ single crystals. The Debye temperatures $\Theta_{\rm D}$ obtained at low~$T$ and for all~$T$ from heat capacity measurements and the Debye temperature $\Theta_{\rm R}$ obtained from fitting electrical resistivity data, respectively, are also listed.}
\begin{ruledtabular}
\begin{tabular}{lcccccccc}

Compound 				& $\gamma_{\rm n} $  	& $\beta $  & $\delta $  &	${\cal D}_C(E_{\rm F})$ 	&	${\cal D}_{\rm band}(E_{\rm F})$ 	&	$\Theta_{\rm D}$ (K)	&  $\Theta_{\rm D}$ (K)		& $\Theta_{\rm R}$ (K) \\	
								& (mJ/mol\,K$^2$)	& (mJ/mol\,K$^4$)	& ($\mu$J/mol\,K$^6$)& (states/eV\,f.u.)  & (states/eV\,f.u.) 		& from low-$T$  			&  from all~$T$			& from $\rho(T)$ \\
\hline
CaPd$_2$As$_2$  					& 6.52(2) 			&  0.463(6) 	& 6.8(4)	&  2.76(1)  & 1.87(1) 					&  	276(1)				& 252(2)  				&  135(3) \\		

SrPd$_2$As$_2$  					& 6.43(3) 			&  0.369(8) 	& 3.7(5) 	&  2.73(2)    & 1.89(1)					&  	298(3)				& 245(3)  				&  170(3) \\

BaPd$_2$As$_2$  					& 4.79(2) 			&  0.638(5) 	& 4.0(3) 	&   2.03(1)    & ---					&  	248(1)				& 227(2)  				&  114(1) \\

\end{tabular}
\end{ruledtabular}
\end{table*}

\begin{figure}
\includegraphics[width=3.3in]{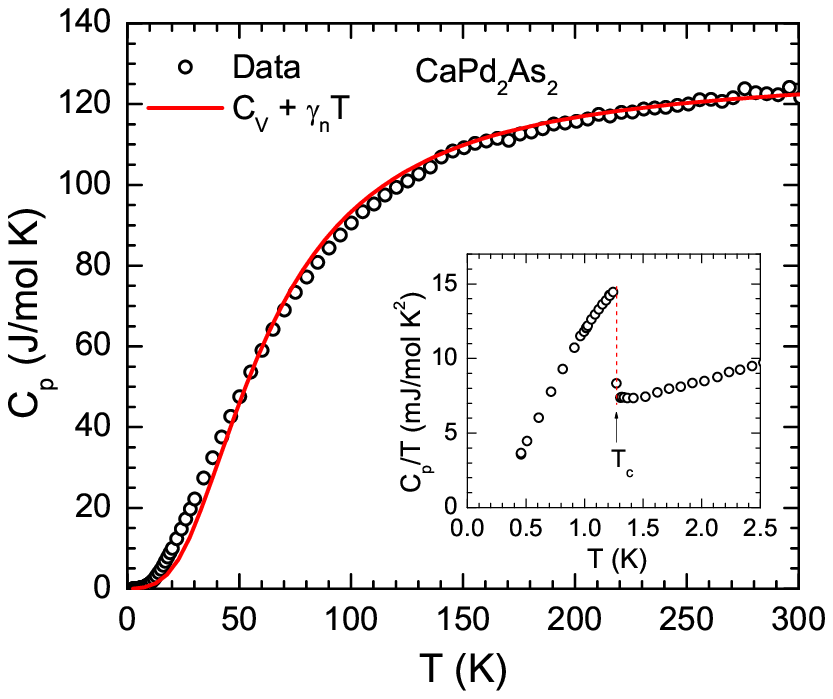}
\caption{(Color online) The heat capacity $C_{\rm p}$ of a CaPd$_2$As$_2$ single crystal [crystal \#2 of Fig.~\ref{fig:CaPd2As2_Rho}(b)] versus temperature~$T$ for 2.5~K~$\leq T \leq$~300~K measured in zero magnetic field. The red solid curve is the fitted sum of the contributions from the Debye lattice heat capacity $C_{\rm V\,Debye}(T)$ and predetermined electronic heat capacity $\gamma_{\rm n} T$ according to Eq.~(\ref{eq:Debye_HC-fit}). Inset: $C_{\rm p}/T$ vs. $T$ for $0.45~{\rm K} \leq T \leq 2.5$~K\@. The dotted red vertical line indicates the $T_{\rm c}$.}
\label{fig:CaPd2As2_HC}
\end{figure}

\begin{figure}
\includegraphics[width=3.3in]{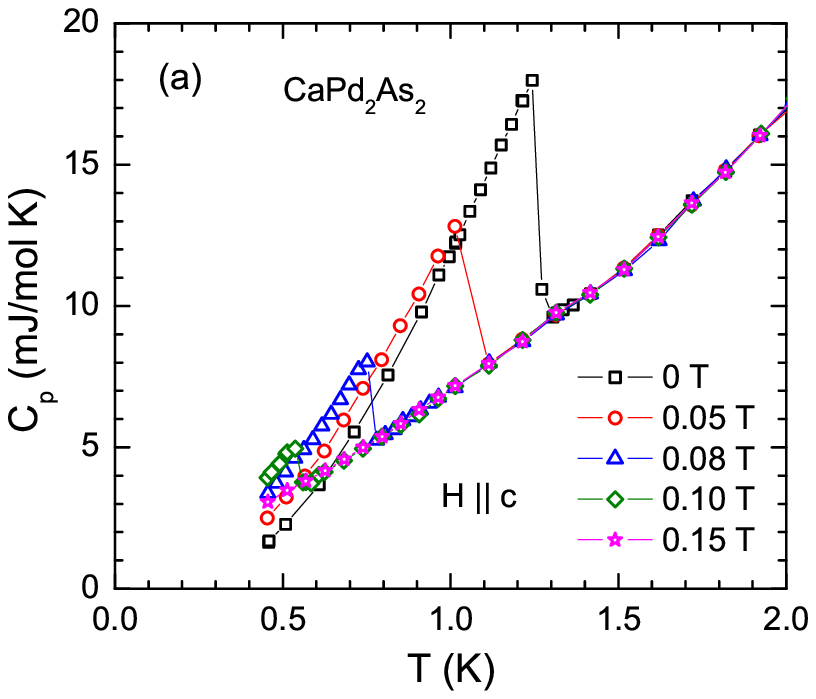}
\includegraphics[width=3.3in]{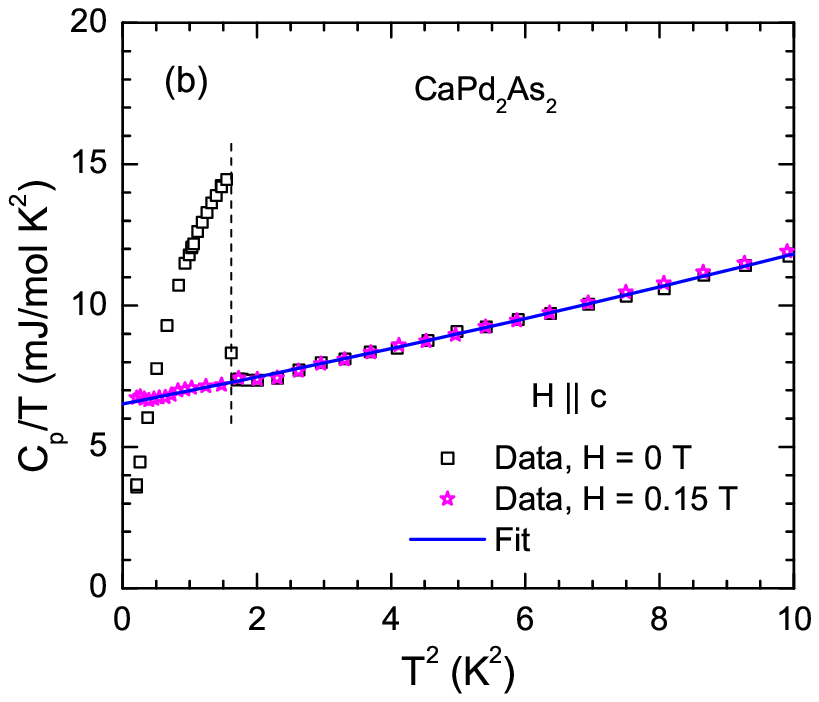}
\caption{(Color online) (a) Heat capacity $C_{\rm p}$ versus temperature $T$ of a CaPd$_2$As$_2$ single crystal [crystal \#2 of Fig.~\ref{fig:CaPd2As2_Rho}(b)] for 0.45~K~$\leq T \leq$~2.0~K measured in different magnetic fields $H$ applied  along the $c$-axis. (b) $C_{\rm p}/T$ versus $T^2$ for 0.45~K~$\leq T \leq$~3.2~K with $H = 0$ and 0.15~T\@. The blue curve is a fit of the $H = 0.15$~T data for $0.45~{\rm K} \leq T \leq 4.5$~K together with the $H = 0$ data for $1.5~{\rm K}\leq T \leq 4.5$~K by Eq.~(\ref{eq:gamma}).}
\label{fig:CaPd2As2_HC_field}
\end{figure}

The $C_{\rm p}(T)$ of a ${\rm CaPd_2As_2}$ crystal is shown in Fig.~\ref{fig:CaPd2As2_HC} for 0.45~K~$\leq T \leq$~300~K\@. As shown in the expanded plot of the low-$T$ data in the inset of Fig.~\ref{fig:CaPd2As2_HC}, a sharp jump is observed in $C_{\rm p}(T)$ due to the occurrence of superconductivity at $T_{\rm c} = 1.27(3)$~K\@. The $C_{\rm p}(T)$ data measured under various $H$ are shown in Fig.~\ref{fig:CaPd2As2_HC_field}(a).  These data show that the $T_{\rm c}$ decreases with increasing $H$ and is suppressed to below 0.45~K by $H = 0.15$~T\@.

Plots of $C_{\rm p}/T$ vs. $T^2$ for $H = 0$ and 0.15~T are shown in Fig.~\ref{fig:CaPd2As2_HC_field}(b).  We analyzed the normal-state $C_{\rm p}(T)$ data at $0.45~{\rm K} \leq T \leq 4.5$~K for $H=0.15$~T and the data at $1.5~{\rm K}\leq T \leq 4.5$~K for $H=0$ together according to
\begin{equation}
\frac{C_{\rm p}(T)}{T}=\gamma_{\rm n} + \beta T^2 + \delta T^4,
\label{eq:gamma}
\end{equation}
where $\gamma_{\rm n}$ is the normal state Sommerfeld electronic heat capacity coefficient, $\beta$ is the Debye $T^3$-law lattice heat capacity coefficient and $\delta T^4$ is a higher-order lattice contribution.  A fit of the normal-state data by Eq.~(\ref{eq:gamma}), shown as the blue curve in Fig.~\ref{fig:CaPd2As2_HC_field}(b), gives $\gamma_{\rm n} = 6.52(2)$~mJ/mol\,K$^2$, $\beta= 0.463(6)$~mJ/mol\,K$^4$ and $\delta = 6.8(4)~\mu$J/mol\,K$^6$. We estimate the Debye temperature $\Theta_{\rm D}$ from $\beta$ using the relation \cite{Kittel2005}
\begin{equation}
\Theta_{\rm D} = \left( \frac{12 \pi^{4} R n}{5 \beta} \right)^{1/3},
 \label{eq:Debye-Temp}
\end{equation}
where $R$ is the molar gas constant and $n=5$ is the number of atoms per formula unit (f.u.), yielding $\Theta_{\rm D}= 276(1)$~K\@.

The measured normal-state $C_{\rm p}(T = 300~{\rm K}) = 123$~J/mol\,K in Fig.~\ref{fig:CaPd2As2_HC} is close to the expected classical Dulong-Petit high-$T$ limiting value $C_{\rm V} = 3nR = 15R$ = 124.7~J/mol\,K at constant volume due to acoustic lattice vibrations. \cite{Kittel2005, Gopal1966}  Our normal-state $C_{\rm p}(T)$ data for the $T$ range 2.5~K~$\leq T \leq$~300~K were fitted by the sum of the above electronic term $\gamma_{\rm n} T$ and the Debye model lattice heat capacity $C_{\rm V\,Debye}(T)$ per mole of atoms due to acoustic lattice vibrations according to
\bse
\label{Eqs:AllTCpFit}
\begin{equation}
C_{\rm p}(T) = \gamma_{\rm n} T + n C_{\rm V\,Debye}(T/\Theta_{\rm D}),
\label{eq:Debye_HC-fit}
\end{equation}
where\cite{Gopal1966}
\begin{equation}
C_{\rm{V\,Debye}}(T/\Theta_{\rm D})=9 R \left( \frac{T}{\Theta_{\rm{D}}} \right)^3 {\int_0^{\Theta_{\rm{D}}/T} \frac{x^4 e^x}{(e^x-1)^2}\,dx}
\label{eq:Debye_HC}
\end{equation}
and
\bea
C_{\rm{V\,Debye}}(T/\Theta_{\rm D}=1) &=& 9 R  {\int_0^1 \frac{x^4 e^x}{(e^x-1)^2}\,dx}\label{Eq:DvTQD}\\*
&\approx& 2.8552\,R.\nonumber
\eea
\ese
The integral in Eq.~(\ref{Eq:DvTQD}) can be obtained in closed form but the expression is cumbersome.  When carrying out the least-squares fit of the experimental $C_{\rm p}(T)$ data by Eqs.~(\ref{Eqs:AllTCpFit}) we used the high-accuracy analytic Pad\'e approximant for $C_{\rm V\,Debye}(T/\Theta_{\rm D})$ in Eq.~(\ref{eq:Debye_HC}) that we formulated in Ref.~\onlinecite{Ryan2012} that greatly simplifies the fit.  The fit was carried out using the fixed value $\gamma_{\rm n} = 6.52$~mJ/mol\,K$^2$ obtained above.  Thus $\Theta_{\rm D}$ is the only adjustable parameter.  The fit yielded $\Theta_{\rm D}= 252(2)$~K and is shown by the red curve in Fig.~\ref{fig:CaPd2As2_HC}. This value of $\Theta_{\rm D}$ is close to but slightly smaller than the value $\Theta_{\rm D}= 276(1)$~K obtained above from analysis of the low-$T$ $C_{\rm p}$ data. Such differences are expected due to the $T$ dependence of $\Theta_{\rm D}$.\cite{Ryan2012, Gopal1966}

We estimate the density of states at the Fermi energy for both spin directions obtained from heat capacity measurements ${\cal D}_{C}(E_{\rm F})$ using the relation \cite{Kittel2005}
\be
\gamma_{\rm n} = \frac{\pi^2 k_{\rm B}^2}{3}\, {\cal D}_{C}(E_{\rm F}).
\label{Eq:gamman}
\ee
This ${\cal D}_{C}(E_{\rm F})$ contains the enhancement from the many-body electron-phonon interaction $\lambda_{\rm {el-ph}}$.  This is also the density of states that enters the BCS equations for $T_{\rm c}$ and for the BCS gap and thermodynamic properties versus $T$\@.  Using the above $\gamma_{\rm n}$ value gives
\be
{\cal D}_{C}(E_{\rm F}) = 2.76~{\rm states/eV\,f.u.}
\ee
The enhancements of the bare (band structure) density of states ${\cal D}_{\rm band}(E_{\rm F})$ and the bare effective mass $m^\ast_{\rm band}$ due to the electron-phonon interaction are\cite{Grimvall1976}
\bse
\label{Eqs:DEFm*lambda}
\bea
{\cal D}_{C}(E_{\rm F}) &=& {\cal D}_{\rm band}(E_{\rm F})(1 + \lambda_{\rm {e-ph}}),\label{eq:DOS}\\*
m^\ast &=& m^\ast_{\rm band}(1+\lambda_{\rm {e-ph}}).\label{Eq:m*}
\eea
\ese

The $\lambda_{\rm {el-ph}}$ value can be estimated from McMillan's theory \cite{McMillan1968} for the electron-phonon mechanism of superconductivity and is related to $\Theta_{\rm D}$ and $T_{\rm c}$ by
\begin{equation}
\lambda_{\rm {el-ph}}= \frac {1.04+\mu^{*} \ln(\Theta_{\rm D}/1.45\,T_{\rm c})} {(1-0.62\mu^{*})\ln(\Theta_{\rm D}/1.45\,T_{\rm c}) - 1.04}.
\label{eq:lambda}
\end{equation}
Here $\mu^{*}$ is the repulsive screened Coulomb parameter having a value often between 0.1 and 0.15 and is usually taken as $\mu^{*} = 0.13$. With this value of $\mu^{*}$ together with $T_{\rm c} = 1.27$~K and $\Theta_{\rm D}= 276$~K as determined above (Table~\ref{tab:HCFitParams}), Eqs.~(\ref{eq:DOS}) and~(\ref{eq:lambda}) yield
\be
\lambda_{\rm {el-ph}} = 0.474,\quad {\cal D}_{\rm band}(E_{\rm F}) = 1.87(1)~{\rm states/eV\,f.u.}
\ee
The relatively small value of $\lambda_{\rm {el-ph}}$ implies weak-coupling superconductivity in ${\rm CaPd_2As_2}$.  

A difference of $\mu^\ast$ from the assumed value of 0.13 would give a different value of the calculated $\lambda_{\rm {e-ph}}$.  For example, from Eq.~(\ref{eq:lambda}) we obtain $\lambda_{\rm {el-ph}} = 0.421$ if $\mu^\ast = 0.10$ and $\lambda_{\rm {el-ph}} = 0.511$ if $\mu^\ast = 0.15$. Such differences would result in corresponding differences in the calculated ${\cal D}_{\rm band}(E_{\rm F})$.  However, our value of ${\cal D}_{\rm band}(E_{\rm F})$ obtained below for the Sr compound using $\mu^\ast = 0.13$ (Table~\ref{tab:HCFitParams}) precisely agrees with the value ${\cal D}_{\rm band}(E_{\rm F}) = 1.93$~states/eV\,f.u.\ for both spin directions obtained from band calculations for this compound in the reference cited in the Note~Added at the end of this paper.  This agreement also indicates that electron correlation effects are not large, consistent with the $sp$-band-like nature of the $A{\rm Pd_2As_2}$ compounds that we deduce from the normal-state data.

The values of ${\cal D}_C(E_{\rm F})$, $m^\ast$ and $\rho_0$ can be used to estimate the the Fermi velocity (speed) $v_{\rm F}$ and the mean free path $\ell$ for conduction carrier scattering at low~$T$\@. In a single-band quasi-free electron Fermi gas model, $v_{\rm F}$ is\cite{Kittel2005}
\bse
\bea
v_{\rm F} &=& \frac {\pi^2 \hbar^3}{{{m^*}^2} V_{\rm f.u.}} {\cal D}_C(E_{\rm F})\label{eq:vF} \\*
 &=& \frac {\pi^2 \hbar^3}{{{m^*}^2} V_{\rm f.u.}} {\cal D}_{\rm band}(E_{\rm F})(1+\lambda_{\rm el-ph}),\label{eq:vF2}
\eea
\ese
where $V_{\rm f.u.} = V_{\rm cell}/2$ from Table~\ref{tab:XRD1} is the volume per formula unit and $\hbar$ is Planck's constant divided by $2\pi$.   Assuming $m^*_{\rm band} = m_{\rm e}$ where $m_{\rm e}$ is the free-electron mass,\cite{Kim2012} Eqs.~(\ref{Eq:m*}) and~(\ref{eq:vF}) yield
\be
v_{\rm F}  = 1.20 \times 10^8~{\rm cm/s}.
\label{Eq:vF1}
\ee
The mean free path $\ell = v_{\rm F}\tau$, where $\tau$ is the mean free scattering time, is obtained from $v_{\rm F}$ using\cite{Kittel2005}
\be
\ell = 3\pi^2 \left(\frac{\hbar}{e^2\rho_0}\right)\left(\frac{ \hbar}{m^\ast v_{\rm F}}\right)^2,
\label{eq:lvF}
\ee
where $\hbar/e^2 = 4108~\Omega$. From Eqs.~(\ref{Eqs:DEFm*lambda}), (\ref{eq:vF}) and~(\ref{eq:lvF}), the value of $m^\ast v_{\rm F}$ and hence of $\ell$ is independent of $\lambda_{\rm el-ph}$.   Using $\rho_0 = 34.6~\mu\Omega$\,cm (for crystal~\#2), $m^*_{\rm band} = m_{\rm e}$ and the value of $v_{\rm F}$ in Eq.~(\ref{Eq:vF1}), Eqs.~(\ref{Eq:m*}) and~(\ref{eq:lvF}) give
\be
\ell = 1.52~{\rm nm},
\label{Eq:mfp1}
\ee
which is only 3.6~in-plane lattice constants (Table~\ref{tab:XRD1}).

The plasma angular frequency $\omega_{\rm p}$ of the conduction carriers can be estimated using the quasi-free-electron single-band relation
\be
n = \frac{1}{3\pi^2}\left(\frac{m^\ast v_{\rm F}}{\hbar}\right)^3,
\ee
yielding\cite{Kittel2005}
\be
\omega_{\rm p}^2 = \frac{4\pi n e^2}{m^*} = \frac{4(m^\ast e)^2(v_{\rm F}/\hbar)^3}{3\pi}.
\label{Eq:omegapDef}
\ee
Using $m^\ast=m_{\rm e}$ and $v_{\rm F}$ from Eq.~(\ref{Eq:vF1}) gives
\be
\omega_{\rm p} = 1.61\times10^{16}~{\rm rad/s}.
\label{Eq:omegap}
\ee
The superconducting London penetration depth in the clean limit at $T=0$, $\lambda_{\rm L}(0)$, is a normal-state property given by\cite{Tinkham1996,Johnston2013}
\be
\lambda_{\rm L}(0) = \frac{c}{\omega_{\rm p}},
\label{Eq:lambda0}
\ee
where $c$ is the speed of light in vacuum.  The value of $\omega_{\rm p}$ in Eq.~(\ref{Eq:omegap}) gives
\be
\lambda_{\rm L}(0) = 1.86\times10^{-6}~{\rm cm} = 18.6~{\rm nm}.
\ee
However, we will see in the following section that CaPd$_2$As$_2$ is in the dirty limit and not in the clean limit that was treated by BCS, where the actual penetration depth $\lambda_{\rm eff}(0)$ is much larger than $\lambda_{\rm L}(0)$.

The normal-state parameters obtained from the above analyses of $C_{\rm p}(T)$ for CaPd$_2$As$_2$ are summarized in Tables~\ref{tab:HCFitParams} and~\ref{tab:SCParams} together with those obtained below for the other two compounds discussed in this paper.  

The Debye temperatures $\Theta_{\rm R}$ in Table~\ref{Tab:RhoFitParams} obtained for the three compounds from analyses of the respective $\rho(T)$ data are also listed in Table~\ref{tab:HCFitParams} for comparison.  The large discrepancies between the values of $\Theta_{\rm D}$ and $\Theta_{\rm R}$ for each compound indicate that the assumptions\cite{Ryan2012} of the BG theory are violated in the $A$Pd$_2$As$_2$ compounds.  The BG theory ignores Umklapp carrier scattering; if the Fermi wave vector $k_{\rm F}$  is significantly smaller than the Debye wave vector $k_{\rm D}$, one might expect $\Theta_{\rm D}$ and $\Theta_{\rm R}$ to be significantly different; and a mechanism such as electron-electron scattering in addition to the electron-phonon scattering assumed in the BG theory could contribute to the $T$ dependence of $\rho$.

\subsubsection{Superconducting State Properties}

\begin{figure}
\includegraphics[width=3.3in]{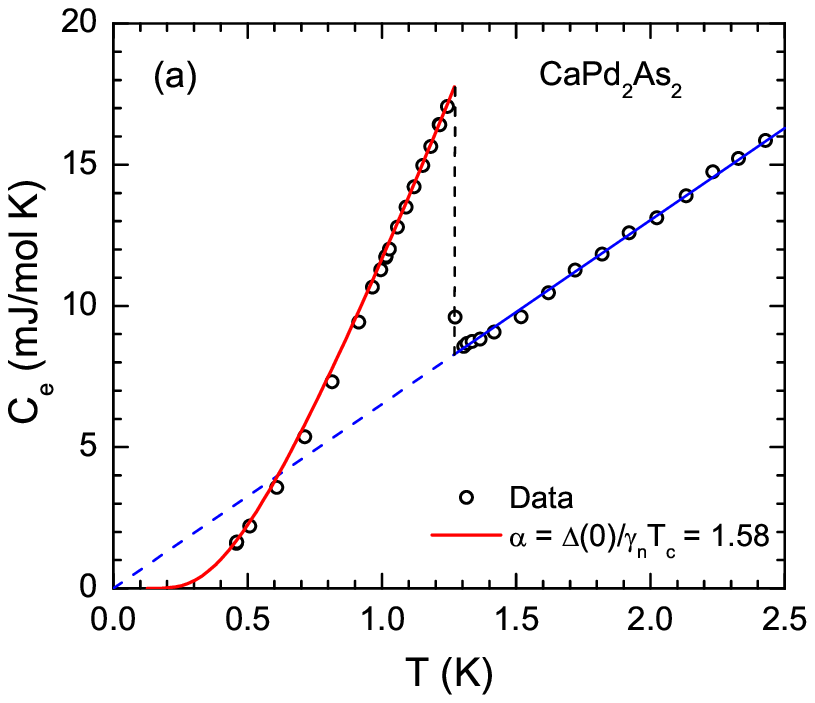}
\includegraphics[width=3.3in]{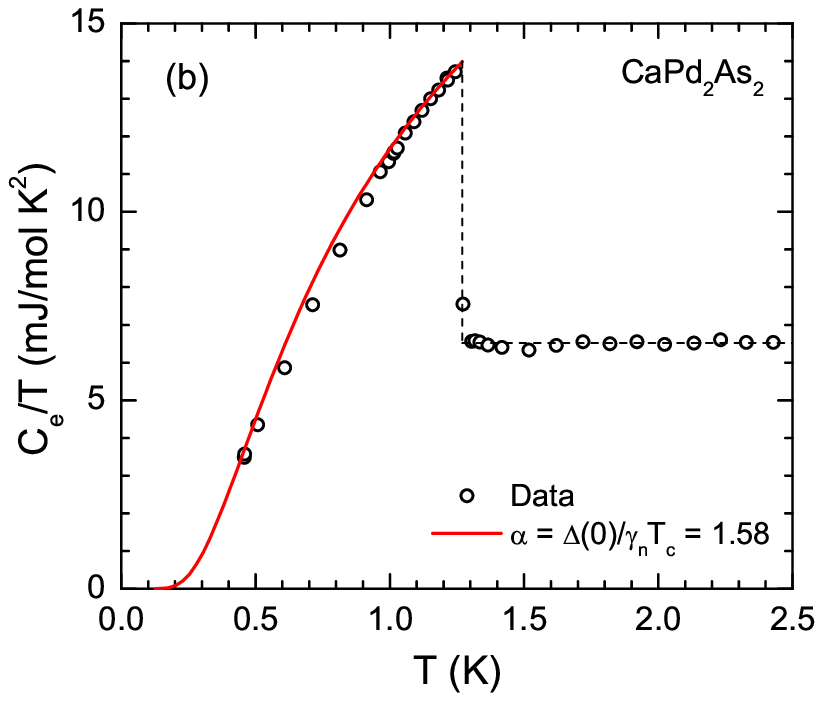}
\caption{(Color online) (a) Electronic contribution $C_{\rm e}$ to the heat capacity versus temperature $T$ of ${\rm CaPd_2As_2}$. (b) $C_{\rm e}/T$ versus $T$\@. The solid red curve in each figure is the theoretical prediction of the $\alpha$-model in Eq.~(\ref{Eq:Ces2}) for $\alpha = 1.58$.  The BCS value is $\alpha_{\rm BCS} \approx 1.764$.  The normal-state heat capacity is $C_{\rm en} = \gamma_{\rm n}T$ with $\gamma_{\rm n} = 6.52~{\rm mJ/mol\,K^2}$.  In (a), this normal-state contribution is the blue straight line and the extrapolation to low~$T$ is the dashed blue line.  In (b), the normal-state behavior is $C_{\rm en}/T = \gamma_{\rm n}$.}
\label{fig:CaPd2As2_HC_el}
\end{figure}

The electronic contribution $C_{\rm e}(T)$ to $C_{\rm p}(T)$ of ${\rm CaPd_2As_2}$ obtained at low~$T$ by subtracting the low-$T$ phonon contribution $\beta T^3 + \delta T^5$ from $C_{\rm p}(T)$ is shown in Fig.~\ref{fig:CaPd2As2_HC_el}(a). The large sharp jump $\Delta C_{\rm e}$ in $C_{\rm e}$ at $T_{\rm c} = 1.27(3)$~K indicates that the superconductivity in ${\rm CaPd_2As_2}$ is a bulk effect, confirmed below.  The vertical heat capacity jump indicated by the vertical dotted line in Fig.~\ref{fig:CaPd2As2_HC_el}(b) yields $\Delta C_{\rm e}/T_{\rm c} = 7.4(2)~{\rm mJ/mol\,K^2}$ and hence $\Delta C_{\rm e} = 9.4(3)$~mJ/mol\,K\@.  Using the normal-state $\gamma_{\rm n} =6.52(2)$~mJ/mol\,K$^2$ from Table~\ref{tab:HCFitParams}, one obtains $\Delta C_{\rm e}/ \gamma_{\rm n} T_{\rm c} = 1.14(3)$ which is much smaller than the BCS value $\Delta C_{\rm e}/ \gamma_{\rm n} T_{\rm c} =1.43$ in the weak-coupling limit.\cite{Tinkham1996,Johnston2013}  Because the bulk superconducting transition in $C_{\rm e}(T)$ is very sharp, we infer that the reduction in $\Delta C_{\rm e}/ \gamma_{\rm n} T_{\rm c}$ from the BCS value is intrinsic.  This reduction can happen if the superconducting gap (order parameter) is anisotropic in wave vector space, either from anisotropy in the gap on a single Fermi surface\cite{Johnston2013} or from multiple bands with distinct Fermi surfaces, each  with a different isotropic or anisotropic gap,\cite{Kogan2009} as discussed in the context of the $\alpha$-model below.

\subsubsection*{The Value of $\alpha$ within the $\alpha$-Model}

For simplicity, we analyze the zero-field superconducting state thermodynamic data for ${\rm CaPd_2As_2}$ within the framework of the so-called $\alpha$-model of the BCS theory of superconductivity,\cite{Padamsee1973, Bardeen1957, Johnston2013} where $\alpha \equiv \Delta(0)/k_{\rm B}T_{\rm c}$. In this model the normalized gap $\tilde{\Delta}(T) \equiv \Delta(T)/\Delta(0)$ is the same as in the BCS theory which is calculated from the BCS gap equation\cite{Bardeen1957, Johnston2013} with $\alpha = \alpha_{\rm BCS}$ where
\be
\alpha_{\rm BCS} = \pi\, e^{-\gamma_{\rm E}}\approx 1.7639
\label{Eq:aBCSDef}
\ee
and $\gamma_{\rm E}\approx 0.5772$ is Euler's constant.  However, for calculating the thermodynamic properties one uses a variable $\alpha$ parameter, which represents an inconsistency in the model, but which allows one to fit superconducting state thermodynamic data that deviate from the BCS predictions.  The superconducting and normal state electronic entropies at $T_{\rm c}$ are the same, as in the BCS model, so the superconducting transition is still second order with no latent heat.  One can interpret the quantitatively determined deviations from the BCS predictions in terms of other models and theories.  For example, values $\alpha>\alpha_{\rm BCS}$ can arise from the presence of strong electron-phonon coupling in contrast to the weak-coupling assumption in the BCS theory,\cite{Carbotte1990} whereas $\alpha<\alpha_{\rm BCS}$ can occur from gap anisotropy in momentum space\cite{Johnston2013} which we therefore assume is responsible for the reduced heat capacity jump in ${\rm CaPd_2As_2}$ compared to the BCS theory prediction.

The value of $\alpha$ is related to the heat capacity jump at $T_{\rm c}$ by\cite{Johnston2013}
\bea
\frac{\Delta C_{\rm e}(T_{\rm c})}{\gamma_{\rm n}T_{\rm c}} &=& \frac{\Delta C_{\rm e}(T_{\rm c})}{\gamma_{\rm n}T_{\rm c}}\bigg|_{\rm BCS}\left(\frac{\alpha}{\alpha_{\rm BCS}}\right)^2\label{Eq:DelCgamTcDelta}\\*
&=& \frac{12}{7\zeta(3)}\left(\frac{\alpha}{\alpha_{\rm BCS}}\right)^2\approx  1.426\left(\frac{\alpha}{\alpha_{\rm BCS}}\right)^2,\nonumber
\eea
where $\zeta(x)$ is the Riemann zeta function.  Inserting our normalized experimental heat capacity jump value into Eq.~(\ref{Eq:DelCgamTcDelta}) gives
\be
\alpha = 1.58(2),
\label{Eq:alphaCaPd2As2}
\ee
which is significantly smaller than the BCS value of~1.764.

The temperature dependence of the superconducting-state electronic heat capacity is calculated within the $\alpha$-model using\cite{Johnston2013}
\bse
\label{Eqs:CesCalcs}
\be
\frac{C_{\rm es}(t)}{\gamma_{\rm n}T_{\rm c}} = \frac{6\alpha^3}{\pi^2t}\int_0^\infty f(1-f)\left(\frac{\tilde{E}^2}{t}-\frac{1}{2}\,\frac{d\tilde{\Delta}^2}{dt}\right)d\tilde{\epsilon},
\label{Eq:Ces2}
\ee
where $t=T/T_{\rm c}$, the normalized normal-state electron energy is $\tilde{\epsilon} = \epsilon/\Delta(0)$, the normalized excited quasiparticle (electron and hole) energy is
\be
\tilde{E} = \frac{E}{\Delta(0)} = \sqrt{\tilde{\epsilon}^2+\tilde{\Delta}^2},
\label{Eq:EbarDef}
\ee
and the Fermi-Dirac distribution function in the dimensionless variables is (with $E_{\rm F}\equiv 0$)
\bea
f\equiv f(\alpha,\tilde{E},t) &=& \frac{1}{e^{\alpha\tilde{E}/t} + 1}\label{Eq:FermiFcnRed}\\*
{\rm with}\ \ \ \frac{E}{k_{\rm B}T} &=&  \frac{\alpha\tilde{E}}{t} .\nonumber
\eea
\ese
The $t$-dependent $\tilde{\Delta}(t)$ and $d\tilde{\Delta}(t)/dt^2$ are calculated as described in Ref.~\onlinecite{Johnston2013}.

The solid red curves in Figs.~\ref{fig:CaPd2As2_HC_el}(a) and~\ref{fig:CaPd2As2_HC_el}(b) are the theoretical predictions for $C_{\rm es}(T)$ and $C_{\rm es}(T)/T$, respectively, calculated from Eq.~(\ref{Eq:Ces2}) using $\alpha = 1.58$ and are seen to be in good agreement with the data. The agreement of the lowest-$T$ data with the theory indicates that there is no residual electronic specific heat for $T\to0$, which in turn indicates that the entire sample is superconducting with a single nodeless $s$-wave gap and with a single $T_{\rm c}$.  Because $\alpha<\alpha_{\rm BCS}\approx 1.764$, we infer that the $s$-wave order parameter is anisotropic in momentum space as discussed above.\cite{Johnston2013}

\begin{figure}
\includegraphics[width=3.3in]{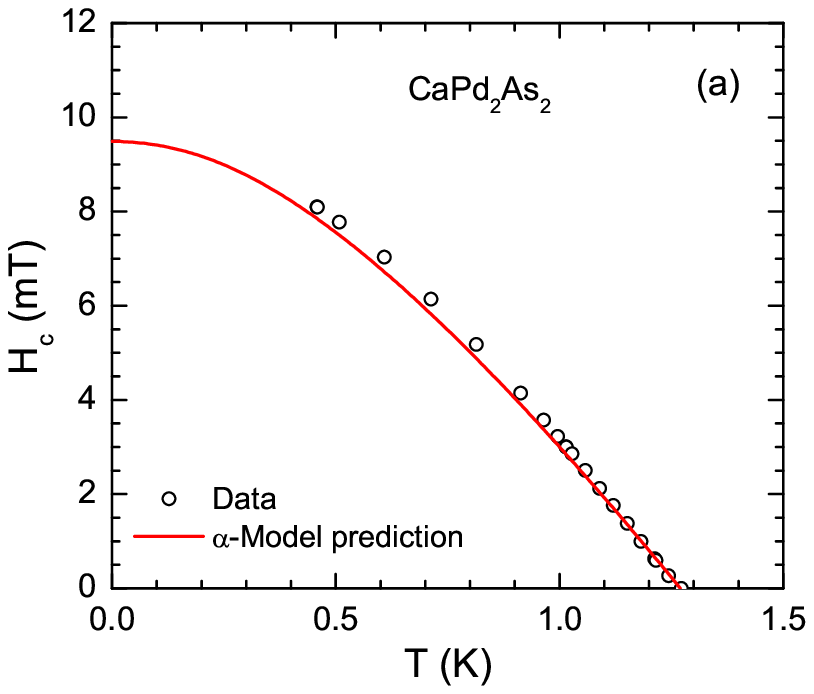}
\includegraphics[width=3.3in]{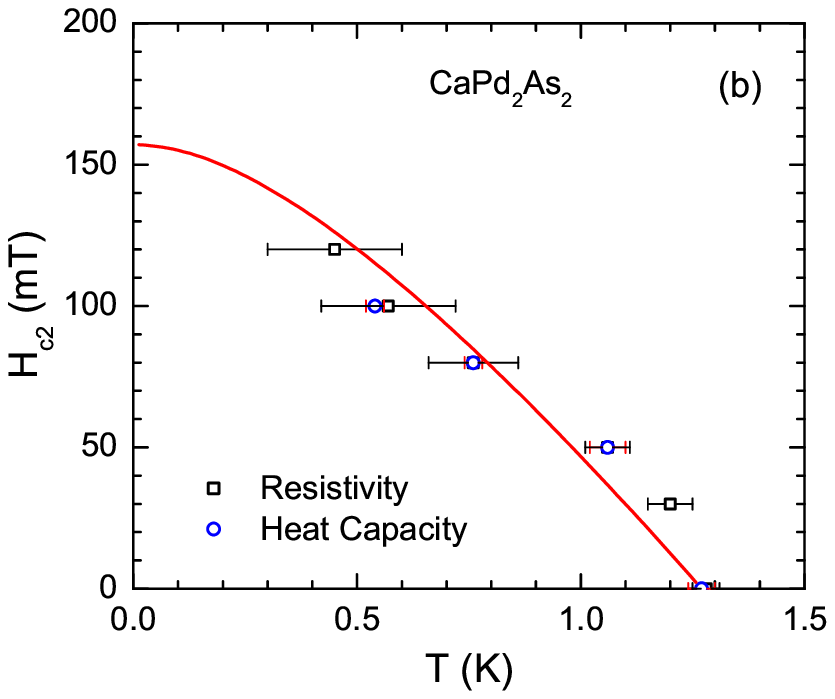}
\caption{(Color online) (a) Thermodynamic critical field $H_{\rm c}$ versus temperature~$T$ obtained for CaPd$_2$As$_2$ from the experimentally derived electronic heat capacity $C_{\rm e}(T)$ data using Eqs.~(\ref{Eqs:HcFromSe}) (open circles) and the theoretical prediction of the $\alpha$-model\cite{Padamsee1973, Johnston2013} using $T_{\rm c} = 1.27~K$, $\alpha=1.58$ and $\gamma_{\rm n} = 6.52~{\rm mJ/mol\,K^2}$ in Eqs.~(\ref{Eq:HcFromGammaTc}) and~(\ref{Eqs:Hc(T)Calc}) (red curve). (b) Upper critical magnetic field $H_{\rm c2}(T)$ of CaPd$_2$As$_2$ determined from the electrical resistivity $\rho(T, H)$ (black open squares) and heat capacity $C_{\rm p}(T, H)$ (blue open circles) data in Figs.~\ref{fig:CaPd2As2_Rho}(b) and \ref{fig:CaPd2As2_HC_field}(a), respectively. The red curve is the prediction for $H_{\rm c2}(T)$  by the WHH theory\cite{WHH1966} in Eqs.~(\ref{Eqs:WHHEqns}) for~$dH_{\rm c2}(T)/dT|_{T=T_{\rm c}} = - 0.18$~T/K, $\alpha_{\rm M}=0.13$, and~$\lambda_{\rm so}=0$.}
\label{fig:CaPd2As2_critical_H}
\end{figure}

\subsubsection*{Thermodynamic Critical Field $H_{\rm c}$}

The experimental thermodynamic critical field $H_{\rm c}$ versus $T$ of a superconductor can be estimated using the zero-field $C_{\rm e}(T)$ data via the electronic entropy difference between the normal ($S_{\rm en})$ and superconducting ($S_{\rm es})$ states per unit volume at  $H=0$ according to the Clausius-Clapeyron-like relation\cite{Tinkham1996, DeGennes1966}
\bse
\label{Eqs:HcFromSe}
\be
S_{\rm en}(T)-S_{\rm es}(T) = -\frac{1}{8\pi}\frac{dH_{\rm c}^{2}(T)}{dT}.
\ee
which, with $S_{\rm e}(T^{\prime}) = \int_0^{T^\prime}[C_{\rm e}(T^{\prime\prime})/T^{\prime\prime})]dT^{\prime\prime}$, yields
\be
H_{\rm c}^{2}(T) = 8\pi\int_{T}^{T_{\rm c}}[S_{\rm en}(T^\prime)-S_{\rm es}(T^\prime)] dT^\prime.
\label{Eq:Hc2FromCp}
\ee
\ese
In cgs units, $H_{\rm c}$ is expressed in units of~Oe, where ${\rm 1~Oe^2 = 1~erg/cm^3}$, so $S_{\rm en}(T^\prime) = \gamma_{\rm nV}T^\prime$ is in units of ${\rm erg/cm^3\,K}$.  The Sommerfeld coefficient $\gamma_{\rm nV}$ in units of ${\rm erg/cm^3\,K^2}$ is calculated from the above Sommerfeld coefficient $\gamma_{\rm n}$ in units of ${\rm mJ/mol~K^2 = 10^4\,erg/mol~K^2}$ according to
\be
\gamma_{\rm nV} = \frac{\gamma_{\rm n}}{V_{\rm M}},
\ee
where $V_{\rm M}$ is the molar volume in units of ${\rm cm^3/mol}$.  For ${\rm CaPd_2As_2}$, our values $\gamma_{\rm n} = {\rm 6.52~mJ/mol\,K^2}$ and $V_{\rm M} = 55.7~{\rm cm^3/mol}$ from Table~\ref{tab:XRD1} give $\gamma_{\rm nV} = 1170~{\rm erg/cm^3\,K^2}$. The experimental $H_{\rm c}(T)$ data obtained from Eqs.~(\ref{Eqs:HcFromSe}) are plotted as open circles in Fig.~\ref{fig:CaPd2As2_critical_H}(a).

In the $\alpha$-model, the thermodynamic critical field at $T=0$ is given by \cite{Johnston2013}
\be
\frac{H_{\rm c}(0)}{\left(\gamma_{\rm nV}T_{\rm c}^2\right)^{1/2}} = \sqrt{\frac{6}{\pi}}\ \alpha \approx 1.382\,\alpha.
\label{Eq:HcFromGammaTc}
\ee
Using our values of $\gamma_{\rm nV}$, $\alpha = 1.58$, and $T_{\rm c} = 1.27$~K gives $H_{\rm c}(0)= 9.5$~mT\@.  The $T$ dependence of $H_{\rm c}$ is calculated from\cite{Johnston2013}
\bse
\label{Eqs:Hc(T)Calc}
\bea
\frac{H_{\rm c}^2(t)}{H_{\rm c}^2(0)} &=& \frac{4\pi^2}{3\alpha^2} \int_t^1 \left[\frac{S_{\rm en}(t^\prime)}{\gamma_{\rm nV}T_{\rm c}}- \frac{S_{\rm es}(t^\prime)}{\gamma_{\rm nV}T_{\rm c}}\right]dt^\prime\nonumber\\*
&=& \frac{4\pi^2}{3\alpha^2} \int_t^1 \left[t^\prime- \frac{S_{\rm es}(t^\prime)}{\gamma_{\rm nV}T_{\rm c}}\right]dt^\prime,
\label{Eq:Hc2}
\eea
where the normal-state electronic entropy is $S_{\rm en}(t^\prime)/\gamma_{\rm nV}T_{\rm c} = t^\prime$, and the superconducting state entropy is calculated from\cite{Johnston2013}
\be
\frac{S_{\rm es}(t)}{\gamma_{\rm nV}T_{\rm c}} = \frac{6\alpha^2}{\pi^2t}\int_0^\infty f(\alpha,\tilde{E},t)\left(\tilde{E} + \frac{\tilde{\epsilon}^2}{\tilde{E}}\right)d\tilde{\epsilon}.
\label{Eq:Se6}
\ee
\ese
The resulting $H_{\rm c}(T)$ calculated for the above values of $H_{\rm c}(0)$, $\alpha$, $\gamma_{\rm nV}$ and~$T_{\rm c}$ is shown by the red curve in Fig.~\ref{fig:CaPd2As2_critical_H}(a). Good agreement is observed between the temperature dependence of the data and the theoretical prediction of the $\alpha$-model, although the calculated zero-temperature value appears to be a bit low.

\subsubsection*{Upper Critical Field $H_{\rm c2}$}

The $\rho(T)$ data in Fig.~\ref{fig:CaPd2As2_Rho}(b) and the $C_{\rm p}(T)$ data in Fig.~\ref{fig:CaPd2As2_HC_field}(a) for ${\rm CaPd_2As_2}$ in fields aligned along the $c$~axis yield the $T$-dependence of the upper critical field $H_{\rm c2}$ shown in Fig.~\ref{fig:CaPd2As2_critical_H}(b).  One sees that $H_{\rm c2}(T\to0)$ is more than an order of magnitude larger than $H_{\rm c}(0) = 9.5$~mT determined above, indicating that ${\rm CaPd_2As_2}$ is a type-II superconductor.  Although the demagnetization factor of the plate-like crystal is large for $H\parallel c$, this has no influence on the present discussion because we only discuss the high-field behavior at the boundary with the normal state, where the magnetization is small.

For a one-band type-II BCS superconductor the orbital critical field $H_{\rm c2}^{\rm Orb}$ at $T=0$ is given by\cite{Hefland1966,WHH1966}
\be
H_{\rm c2}^{\rm Orb}(0) = - A\,T_{\rm c} \frac{dH_{\rm c2}(T)}{dT}\bigg|_{T=T_{\rm c}},
\label{Eq:Hc2Orb}
\ee
where $A = 0.73$ and~0.69 in the clean and dirty limits, respectively.  From Fig.~\ref{fig:CaPd2As2_critical_H}(b), $dH_{\rm c2}(T)/dT|_{T=T_{\rm c}} = - 0.22(6)$~T/K as determined from the $\rho(T)$ data for $0.8  < T/T_{\rm c} < 1.0$. Thus for CaPd$_2$As$_2$ we estimate $H_{\rm c2}^{\rm Orb}(0) = 0.20(6)$~T in the clean limit and $H_{\rm c2}^{\rm Orb}(0) = 0.19(5)$~T in the dirty limit.

The Pauli-limiting upper critical field at $T=0$, $H_{\rm P}(0)$, is the field at which the magnetic field energy of the current carriers is equal to the superconducting condensation energy.\cite{Clogston1962,Chandrasekhar1962}  Within the $\alpha$-model, this is given for spectroscopic splitting factor $g=2$ by
\be
\frac{\mu_{\rm B}H_{\rm P}(0)}{k_{\rm B}T_{\rm c}} = \frac{\alpha}{\sqrt{2}} \approx 1.2473 \left(\frac{\alpha}{\alpha_{\rm BCS}}\right),
\ee
where $\mu_{\rm B}$ is the Bohr magneton and $\alpha_{\rm BCS}$ is given in Eq.~(\ref{Eq:aBCSDef}), yielding
\be
H_{\rm P}(0)[{\rm T}] = 1.86\,T_{\rm c}[{\rm K}]\left(\frac{\alpha}{\alpha_{\rm BCS}}\right).
\label{Hp(0)}
\ee
Taking $T_{\rm c} = 1.27$~K, $\alpha = 1.58$ from Eq.~(\ref{Eq:alphaCaPd2As2}) and $\alpha_{\rm BCS} = 1.7639$ from Eq.~(\ref{Eq:aBCSDef}) gives $H_{\rm P}(0) = 2.12$~T\@.  Since the measured $H_{\rm c2}(0) \sim 0.15~{\rm T}\sim 0.07\, H_{\rm P}(0)$ [see Fig.~\ref{fig:CaPd2As2_critical_H}(b)], the effects of Pauli limiting on $H_{\rm c2}$ should be minimal.  The Maki parameter $\alpha_{\rm M}$ expresses the relative magnitudes of the orbital and Pauli limiting $H_{\rm c2}$ values as\cite{Maki1966}
\be
\alpha_{\rm M} = \sqrt{2}\ \frac{H_{\rm c2}^{\rm Orb}(0)}{H_{\rm P}(0)},
\label{Eq:alphaMDef}
\ee
which gives $\alpha_{\rm M} = 0.13$ for ${\rm CaPd_2As_2}$.

To include the influence of Pauli limiting and spin-orbit scattering of quasiparticles on $H_{\rm c2}$, we analyzed the $H_{\rm c2}(T)$ data within the Werthamer, Helfand and Hohenberg (WHH) theory for a one-band type-II dirty-limit superconductor which calculates $H_{\rm c2}$ in terms of the orbital, spin-orbit scattering and Pauli spin paramagnetism contributions in dimensionless variables as\cite{WHH1966}
\bse
\label{Eqs:WHHEqns}
\begin{eqnarray}
\ln\frac{1}{t} & = &\sum_{\nu=-\infty}^\infty \Bigg\{ \frac{1}{|2\nu+1|} - \bigg[ |2\nu+1| + \frac{\bar{h}}{t}  \label{eq:WHH} \\
 && \hspace{0.6in}+\ \frac{(\alpha_{\rm M}\bar{h}/t)^2}{|2\nu+1|+(\bar{h} +\lambda_{\rm so})/t}\bigg] ^{-1}  \Bigg\},
\nonumber
\end{eqnarray}
where $t = T/T_{\rm c}$, $\lambda_{\rm so}$ is the spin-orbit scattering parameter and
\be
\bar{h} = - \left(\frac{4}{\pi^2}\right) \frac{H_{\rm c2}(T)/T_{\rm c}}{dH_{\rm c2}(T)/dT|_{T=T_{\rm c}}} .
\ee
\ese

WHH state that the applicability of their theory to a specific superconductor can be tested by comparing the value of $\alpha_{\rm M}$ calculated from Eq.~(\ref{Eq:alphaMDef}) with the value obtained from their alternative expression\cite{WHH1966}
\be
\alpha_{\rm M} = -0.528\ \frac{dH_{\rm c2}(T)}{dT}\Big|_{T=T_{\rm c}},
\label{Eq:alphaMWHH}
\ee
where the derivative is in units of T/K\@.  Using our value
\be
\frac{dH_{\rm c2}(T)}{dT}\Big|_{T=T_{\rm c}} = - 0.22(6)~{\rm \frac{T}{K}}
\label{Eq:dHc2dTCa}
\ee
obtained from Fig.~\ref{fig:CaPd2As2_critical_H}(b), Eq.~(\ref{Eq:alphaMWHH}) gives $\alpha_{\rm M} = 0.12(3)$, which is indeed in agreement with the above estimate $\alpha_{\rm M} = 0.13$, thus indicating that the WHH theory is  appropriate for fitting our $H_{\rm c2}(T)$ data.

Reasonable agreement of our $H_{\rm c2}(T)$ data with the WHH theory prediction for $\alpha_{\rm M} = 0.13$ was obtained from Eqs.~(\ref{Eqs:WHHEqns}) with $\lambda_{\rm so} = 0$ and $dH_{\rm c2}(T)/dT|_{T=T_{\rm c}} = - 0.18$~T/K as shown in Fig.~\ref{fig:CaPd2As2_critical_H}(b), where $H_{\rm c2}(0) = 157$~mT\@.  The same value of $dH_{\rm c2}(T)/dT|_{T=T_{\rm c}}$ is obtained from the dirty-limit relation \cite{Orlando1979} $dH_{\rm c2}(T)/dT|_{T=T_{\rm c}} = 4.48 \times 10^{4}\, \gamma_{\rm nV} \rho_{0}$, where $\rho$ is in units of $\Omega$\,cm, indicating that CaPd$_2$As$_2$ is in the dirty limit as further documented below.

\subsubsection*{Ginzburg-Landau Parameter}

The Ginzburg-Landau parameter $\kappa_{\rm GL}$ can be estimated from the relation \cite{Tinkham1996}
\be
\kappa_{\rm GL} = \frac{H_{\rm c2}}{\sqrt{2}\,H_{\rm c}}.
\label{Eq:KappaGL}
\ee
Using the $T\to0$ values $H_{\rm c2} = 157$~mT and $H_{\rm c} = 9.5$~mT gives $\kappa_{\rm GL} = 11.7\gg 1/\sqrt{2}$, characterizing CaPd$_2$As$_2$ as a type-II superconductor.  An estimate of the lower critical field $H_{\rm c1}$ is then obtained from\cite{Tinkham1996}
\be
H_{\rm c1} = H_{\rm c} \frac{\ln \kappa_{\rm GL}}{\sqrt{2}\,\kappa_{\rm GL}},
\label{Eq:Hc1fromKappaGL}
\ee
which for $H_{\rm c}(0) = 9.5$~mT and $\kappa_{\rm GL} = 11.7$ gives $H_{\rm c1}(0) = 1.4$~mT\@.  Then using the above values and the expression applicable to the high-$\kappa_{\rm GL}$ limit\cite{Tinkham1996}
\be
\lambda_{\rm eff}^2 = \frac{\Phi_0 H_{\rm c2}}{4\pi H_{\rm c}^2},
\ee
where $\Phi_0 = 2.07 \times 10^{-7}$~G\,cm$^2$ is the flux quantum, yields the $T\to0$ effective magnetic penetration depth $\lambda_{\rm eff} = 530$~nm in the notation of Tinkham.

The Ginzburg-Landau coherence length at $T=0$, $\xi(0)$, can be estimated from \cite{Tinkham1996, DeGennes1966}
\be
H_{\rm c2}(0) = \frac{\Phi_0}{2\pi \xi(0)^2}.
\label{Eq:xiFromHc2}
\ee
Using $H_{\rm c2} = 157$~mT this gives
\be
\xi(0) = 45.8~{\rm nm}.
\label{Eq:xi0Val}
\ee
This $\xi(0)$ is much larger than the mean free path $\ell = 1.52$~nm in Eq.~(\ref{Eq:mfp1}) which indicates that CaPd$_2$As$_2$ is in the dirty limit.

In the dirty limit the penetration depth at $T\to0$ is given by\cite{Tinkham1996}
\be
\lambda_{\rm eff}(0) = \lambda_{\rm L}(0)\sqrt{1+\frac{\xi_0}{\ell}}\qquad {\rm (dirty\ limit)}.
\label{eq:lambda_eff}
\ee
The $T$ dependence of $\xi$ is given by \cite{Tinkham1996}
\be
\frac{\xi(T)}{\xi_0} =  \frac{\pi}{2\sqrt{3}}\frac{H_{\rm c}(0)}{H_{\rm c}(T)} \frac{\lambda_{\rm L}(0)}{\lambda_{\rm eff}(T)},
\label{eq:xi0l}
\ee
yielding the relationship between the zero-temperature Ginzburg-Landau $\xi(0)$ and the BCS $\xi_0$ as
\be
\frac{\xi(0)}{\xi_0} =  \frac{\pi}{2\sqrt{3}} \frac{\lambda_{\rm L}(0)}{\lambda_{\rm eff}(0)}.
\label{eq:xi0l2}
\ee
Combining Eqs.~(\ref{eq:lambda_eff}) and~(\ref{eq:xi0l2}) gives
\be
\frac{\xi(0)}{\xi_0} = \frac{\pi}{2\sqrt{3\left(1+\frac{\xi_0}{\ell}\right)}}\qquad {\rm (dirty\ limit)}.
\label{Eq:xi0xi00}
\ee
Substituting the above values for $\xi(0)$ in Eq.~(\ref{Eq:xi0Val}) and $\ell$ in Eq.~(\ref{Eq:mfp1}) into Eq.~(\ref{Eq:xi0xi00}) and solving for $\xi_0$ gives $\xi_0 = 1690$~nm.  Then using $\lambda_{\rm L}(0) = 18.6$~nm, $\xi_0 = 1690$~nm and $\ell = 1.52$~nm, Eq.~(\ref{eq:lambda_eff}) gives $\lambda_{\rm eff}(0) = 620$~nm in the dirty limit, in agreement with the above estimate, which is of the same order as the value of 210(60)~nm obtained from the penetration depth measurements in Sec.~\ref{Sec:LondonPD} below.

In the absence of impurity scattering, the Ginzburg-Landau parameter at $T=0$ would have been $\kappa_{\rm GL} = \lambda_{\rm L}^{\rm calc}(0)/\xi_0 = 18.6$~nm/1690~nm $\approx 0.01 \ll 1/\sqrt{2}$ which would have been in the extreme type-I limit instead of in the type-II regime.  A similar situation was found for ${\rm SrPd_2Ge_2}$.\cite{Kim2012}

The BCS coherence length is related to the Fermi velocity within the $\alpha$-model by\cite{Johnston2013}
\be
\xi_0 = \frac{\hbar v_{\rm F}}{\pi \Delta(0)} = \left(\frac{1}{\pi \alpha}\right)\frac{\hbar v_{\rm F}}{k_{\rm B} T_{\rm c}}.
\label{eq:xivF}
\ee
This allows an additional estimate of the Fermi velocity which for $\alpha=1.58$ and $\xi_0 = 1690$~nm yields $v_{\rm F} = 1.40 \times 10^8$~cm/s. This value of $v_{\rm F}$ is close to the value $v_{\rm F}  = 1.20 \times 10^8~{\rm cm/s}$ in Eq.~(\ref{Eq:vF1}) obtained from ${\cal D}_C(E_{\rm F})$ according to Eq.~(\ref{eq:vF}), indicating the self-consistency of our modeling.

A summary of the measured and derived superconducting state parameters for CaPd$_2$As$_2$ is given in Table~\ref{tab:SCParams}.

\begin{table}
\caption{\label{tab:SCParams} Measured and derived superconducting and relevant normal state parameters for CaPd$_2$As$_2$.  $T_{\rm c}$: bulk superconducting transition temperature; $\gamma_{\rm n}$: observed Sommerfeld coefficient of the linear term in the low-$T$ normal-state heat capacity; $\lambda_{\rm el-ph}$: electron-phonon coupling constant; $\Delta C_{\rm e}$: heat capacity jump at $T_{\rm c}$; $\alpha = \Delta(0)/k_{\rm B}T_{\rm c}$; $\Delta$: superconducting order parameter; $\alpha_{\rm M}$: Maki parameter; $H_{\rm c}$, $H_{\rm P}$, $H_{\rm c1}$, $H_{\rm c2}^{\rm Orb}$, $H_{\rm c2}$: thermodynamic, Paul limiting upper critical, lower critical, orbital upper critical, and fitted upper critical magnetic fields, respectively; $\kappa_{\rm GL}$: Ginzburg-Landau parameter; $\xi$: Ginzburg-Landau coherence length; $\xi_0$: BCS superconducting coherence length; $\ell$: electronic mean-free path at low~$T$; $\omega_{\rm p}$: angular plasma frequency; $\lambda_{\rm L}$: London penetration depth; $\lambda_{\rm eff}$: magnetic penetration depth. The value of $\lambda_{\rm eff}^{\rm obs}(0)$ is determined from the magnetic penetration depth measurements.}
\begin{ruledtabular}
\begin{tabular}{lcc}
CaPd$_2$As$_2$ property & value\\
\hline
$T_{\rm c}$ (K)                                         & 1.27(3)    \\
$\gamma_{\rm n}$ (mJ/mol\,K$^{2}$)                      & 6.52(2) \\
$\lambda_{\rm el-ph}$ assuming $\mu^\ast=0.13$           & 0.474       \\
$\Delta C_{\rm e}$ (mJ/mol\,K)                          & 9.4(3)  \\
$\Delta C_{\rm e}/\gamma_{\rm n} T_{\rm c}$             & 1.14(3)  \\
$\alpha$ (from $\Delta C_{\rm e}/\gamma_{\rm n} T_{\rm c}$) & 1.58(2)  \\
$\Delta(0)/k_{\rm B}$~(K) (observed)				& 2.02(14)	\\
$\alpha_{\rm M}$                                        & 0.13       \\
$H_{\rm c}(T=0)$ (mT)                                   & 9.5    \\
$H_{\rm P}$ (T) 									& 2.12 \\
$H_{\rm c1}(T=0)$ (mT)                                  & 1.4      \\
$H_{\rm c2}^{\rm Orb}(T=0)$~(dirty limit) (T)           & 0.19(5) \\
$H_{\rm c2}(T=0)$ (mT)                                  & 157  \\
$\kappa_{\rm GL}$                                       & 11.7   \\
$\xi(T=0)$ (nm)                                         & 45.8        \\
$\xi_0$ (nm)                                            & 1690        \\
$\ell~(m^\ast=m_{\rm e})$ (nm)                          & 1.52       \\
$\omega_{\rm p}~(m^\ast=m_{\rm e})~(10^{16}~{\rm rad/s})$ & 1.61 \\
$\lambda_{\rm L}^{\rm calc}(0)$~(clean limit) (nm)      & 18.6     \\
$\lambda_{\rm eff}^{\rm calc}(0)$~(dirty limit) (nm)    & 530--620  \\
$\lambda_{\rm eff}^{\rm obs}(0)$ (nm)                     & 210(60)        \\
\end{tabular}
\end{ruledtabular}
\end{table}

\subsection{\label{Sec:CaPd2As2_M(H,T)} Magnetization and Magnetic Susceptibility}

\begin{figure}[t]
\includegraphics[width=3.3in]{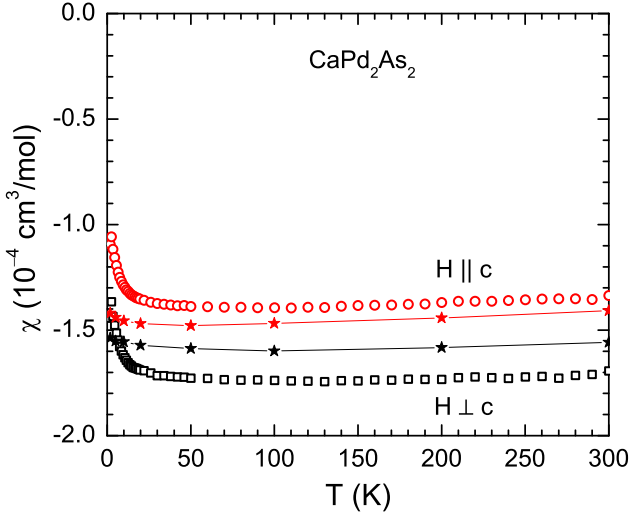}
\caption{ (Color online) Zero-field-cooled magnetic susceptibility $\chi$ of a ${\rm CaPd_2As_2}$ single crystal versus temperature $T$ measured in a magnetic field $H = 3.0$~T applied along the $c$-axis ($\chi_c,\ H \parallel c$) and in the $ab$-plane ($\chi_{ab},\ H \perp  c$). The filled stars represent the intrinsic $\chi$ obtained from fitting $M(H)$ isotherm data in the Appendix by Eq.~(\ref{eq:MH_fit}) which are more accurate that the temperature-scanned data at fixed field. The lines joining the stars are guides to the eye.}
\label{fig:MT_CaPd2As2}
\end{figure}

Zero-field-cooled (ZFC) $\chi(T) \equiv M(T)/H$ data for a ${\rm CaPd_2As_2}$ crystal versus $T$ in a magnetic field $H = 3.0$~T applied along the $c$-axis ($\chi_c,\ H \parallel c$) and in the $ab$-plane ($\chi_{ab},\ H \perp c$) are shown in Fig.~\ref{fig:MT_CaPd2As2}. The data for both directions of $H$ are strongly diamagnetic and nearly independent of $T$ except for the Curie-like upturns at low-$T$ which were found from analysis of $M(H)$ isotherms in the Appendix to be due to the presence of a small amount of saturable paramagnetic (PM) impurities. The intrinsic $\chi(T)$ values at several temperatures obtained from the analysis of the $M(H)$ isotherms are shown by filled stars in Fig.~\ref{fig:MT_CaPd2As2}, which are more accurate than the $\chi(T) = M(T)/H$ data. The $\chi$ is anisotropic with $\chi_{c} > \chi_{ab}$ over the entire $T$ range.

The $\chi$ anisotropy in ${\rm CaPd_2As_2}$ is different from the  anisotropy usually observed in doped and undoped FeAs-based ${\rm ThCr_2Si_2}$-structure compounds, where $\chi_{ab} > \chi_{c}$ such as in BaFe$_2$As$_2$.\cite{Wang2009,Johnston2010}   The powder and temperature average of the intrinsic $\chi$ obtained from fitting the $M(H)$ isotherm data in the Appendix over the $T$ range 20 to 300~K is $\langle\chi\rangle = [2 \langle\chi_{ab}\rangle + \langle\chi_{c}\rangle]/3 = -1.5\times 10^{-4}$~cm$^3$/mol.

The diamagnetic susceptibilities in Fig.~\ref{fig:MT_CaPd2As2} seem to be rather large.  For the purpose of comparing the diamagnetic susceptibilities of different materials, a comparison of the thermodynamic dimensionless magnetic susceptibility per unit volume $\chi_{\rm V}$ (volume susceptibility) is most appropriate.  Using the molar volume $V_{\rm M}=55.7~{\rm cm^3/mol}$ for ${\rm CaPd_2As_2}$ in Table~\ref{tab:XRD1} and $\chi_{\rm ab}(100~{\rm K}) = -1.6\times10^{-4}~{\rm cm^3/mol}$ from Fig.~\ref{fig:MT_CaPd2As2} (black star), one obtains $\chi_{{\rm V}\,ab} = \chi_{ab}/V_{\rm M}$ in the $ab$~plane as
\be
\chi_{{\rm V}ab}(100~{\rm K}) = -0.29\times10^{-5}\quad {\rm(CaPd_2As_2)}.
\label{Eq:chiVCaPd2As2}
\ee
This value can be compared with the respective values for elemental Bi and C (graphite) that are well-known for their exceptionally strong diamagnetism.  Bi has a rhombohedral crystal structure with a mass density $\rho_{\rm m} = 9.8~{\rm g/cm^3}$ and a gram susceptibility in the hexagonal $ab$~plane $\chi_{{\rm g}\,ab}(100~{\rm K) = -1.9\times10^{-6}~cm^3/g}$.\cite{Otake1980}  The volume susceptibility $\chi_{\rm V} = \rho_{\rm m}\chi_{\rm g}$ is then
\be
\chi_{{\rm V}ab}(100~{\rm K}) = -1.9\times10^{-5}\quad {\rm(Bi)}.
\ee
On the other hand, using $\rho_{\rm m} =  2.27$~g/cm$^3$ for hexagonal highly-oriented pyrolytic graphite (HOPG) and $\chi_{{\rm g}\,c}(100~{\rm K) = -2.3\times10^{-5}~cm^3/g}$,\cite{Heremans1994} one obtains
\be
\chi_{{\rm V}c}(100~{\rm K}) = -5.2\times10^{-5}\quad {\rm (C,\ HOPG)}.
\ee
These values for Bi and graphite are about 6.5 and 18 times more diamagnetic than the value for ${\rm CaPd_2As_2}$ in Eq.~(\ref{Eq:chiVCaPd2As2}), respectively.  However, these three values are all much less diamagnetic than the value $\chi_{\rm V} = -1/4\pi \approx -0.0796$ for the diamagnetic susceptibility of a superconductor at low fields with zero demagnetization factor due to complete exclusion of the magnetic induction from the interior (except within a magnetic field penetration depth of the surface).

The different contributions to the intrinsic $\chi $ are
\begin{equation}
\chi=\chi_{\rm core}+\chi_{\rm VV}+\chi_{\rm L} + \chi_{\rm P},
\label{eq:chi}
\end{equation}
where the first three terms are orbital susceptibilities and the last term is the Pauli spin susceptibility of the conduction carriers.  $\chi_{\rm core}$ is the isotropic diamagnetic susceptibility of localized core electrons, $\chi_{\rm VV}$ is the generally anisotropic paramagnetic Van Vleck susceptibility, and $\chi_{\rm L}$ is the generally isotropic Landau diamagnetic susceptibility of the conduction carriers.

The $\chi_{\rm core}$ is estimated using atomic diamagnetic susceptibilities\cite{Mendelsohn1970} which gives $\chi_{\rm {core}}$ = $-$1.78 $\times$ 10$^{-4}$ cm$^3$/mol. The $\chi_{\rm P}$ is related to ${\cal D}(E_{\rm F})$ by \cite{Ashcroft1976,Johnston2010}
\begin{equation}
\chi_{\rm {P}} = \frac{g^2}{4} \mu_{\rm B}^2 {\cal D}_{\rm band}(E_{\rm F}),
\label{eq:Chi-Pauli}
\end{equation}
where we assume that there are no many-body enhancements to $\chi_{\rm {P}}$.  Then using $g=2$ and ${\cal D}(E_{\rm F}) = 1.87$~states/eV\,f.u.\ for both spin directions, from Table~\ref{tab:HCFitParams} we obtain $\chi_{\rm {P}} = 6.0 \times 10^{-5}$~cm$^3$/mol. The $\chi_{\rm L}$ is related to $\chi_{\rm P}$ by\cite{Ashcroft1976, Elliott1998}
\begin{equation}
 \chi_{\rm {L}} = - \frac{1}{3} \left( \frac {m_{\rm e}}{m^*_{\rm band}} \right)^2 \chi_{\rm {P}},
 \label{eq:Chi-Landau}
\end{equation}
where we assume that $\chi_{\rm {L}}$ is not enhanced by the electron-phonon interaction. Assuming $m_{\rm band}^* = m_{\rm e}$ we obtain $\chi_{\rm {L}} = -2.0 \times 10^{-5}$~cm$^3$/mol from the above value of $\chi_{\rm P}$. Then $\chi_{\rm VV}$ is obtained by subtracting these three contributions from the measured $\chi$ according to Eq.~(\ref{eq:chi}).  The four contributions to the intrinsic $\chi$ are summarized in Table~\ref{tab:Chi_Contr}, along with corresponding values for ${\rm SrPd_2As_2}$ and ${\rm BaPd_2As_2}$ determined below.

It is seen in Table~\ref{tab:Chi_Contr} that the inferred value of $\chi_{\rm VV}$ for ${\rm CaPd_2As_2}$ is negative, which is unphysical.  The reason for this error is not clear.  The most likely source of the negative $\chi_{\rm VV}$ value is a small error in correcting the total measured magnetic moment for the sample holder contribution, which was up to 40\% of the total measured moment.  Thus the uncertainty in the measured $\chi$ values is of order 10\% as discussed in Sec.~\ref{ExpDetails}.  In particular, the negative $\chi_{\rm VV}$ is about 8\% of the measured moment, and an error of only 4\% in the sample holder correction could cause the derived $\chi_{\rm VV}$ to be negative.  

\begin{table}
\caption{\label{tab:Chi_Contr} Estimated contributions to the intrinsic angle- and temperature-averaged magnetic susceptibilities $\langle \chi \rangle$ of $A$Pd$_2$As$_2$ ($A$ = Ba, Ca, Sr) crystals.  Here $\chi_{\rm P}$ is the Pauli spin susceptibility of the conduction carriers, and the orbital susceptibility contributions are the  diamagnetism $\chi_{\rm core}$ of the atomic electron cores, the Landau diamagnetism $\chi_{\rm L}$ of the conduction carriers and the Van Vleck paramagnetism $\chi_{\rm VV}$.  All susceptibilities are in units of 10$^{-5}$\,cm$^3$/mol.  Possible reasons for the unphysical negative value of $\langle\chi_{\rm VV}\rangle$ for CaPd$_2$As$_2$ are discussed in the text.}
\begin{ruledtabular}
\begin{tabular}{lccccr}

Compound & $\langle \chi \rangle$ & $\chi_{\rm core}$  &	 $\chi_{\rm {P}}$ 	&	$\chi_{\rm L}$	&	$\langle\chi_{\rm VV}\rangle$ \\	
\hline
CaPd$_2$As$_2$   & $-$15.4  &  $-$17.8 & 6.0 & $-$2.0 & $-1.6$ \\		
SrPd$_2$As$_2$   & $-$3.9  &  $-$19.3 & 6.0 & $-$2.0 & 11.4 \\	
BaPd$_2$As$_2$   & $-$12.4  &  $-$21.6 & 6.6 & $-$2.2 & 4.8 \\	
\end{tabular}
\end{ruledtabular}
\end{table}

\section{\label{SrPd2As2} Physical Properties of S\lowercase{r}P\lowercase{d}$_2$A\lowercase{s}$_2$ Crystals}

\subsection{\label{Sec:SrPd2As2_Rho} Electrical Resistivity}

\begin{figure}[t]
\includegraphics[width=3.3in]{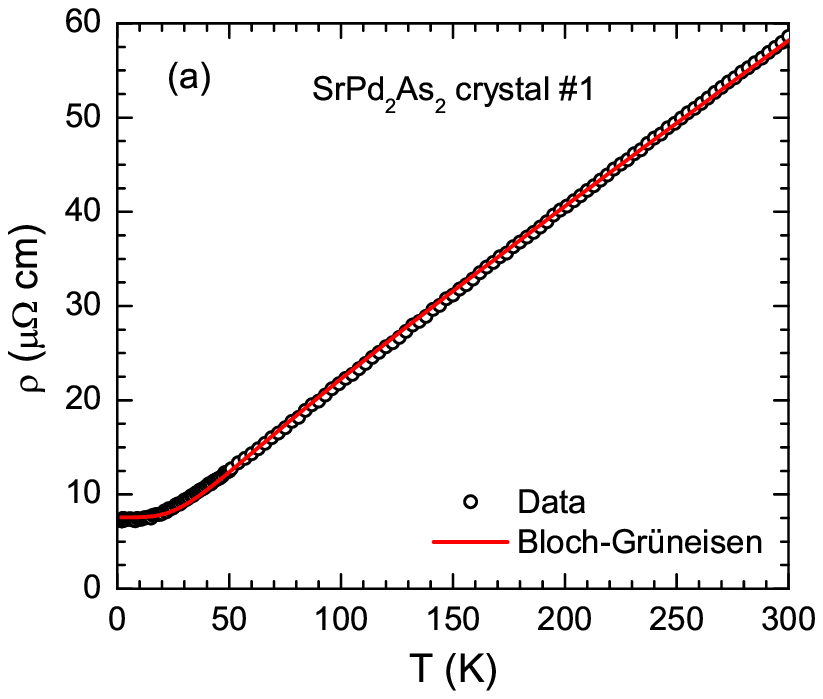}
\includegraphics[width=3.3in]{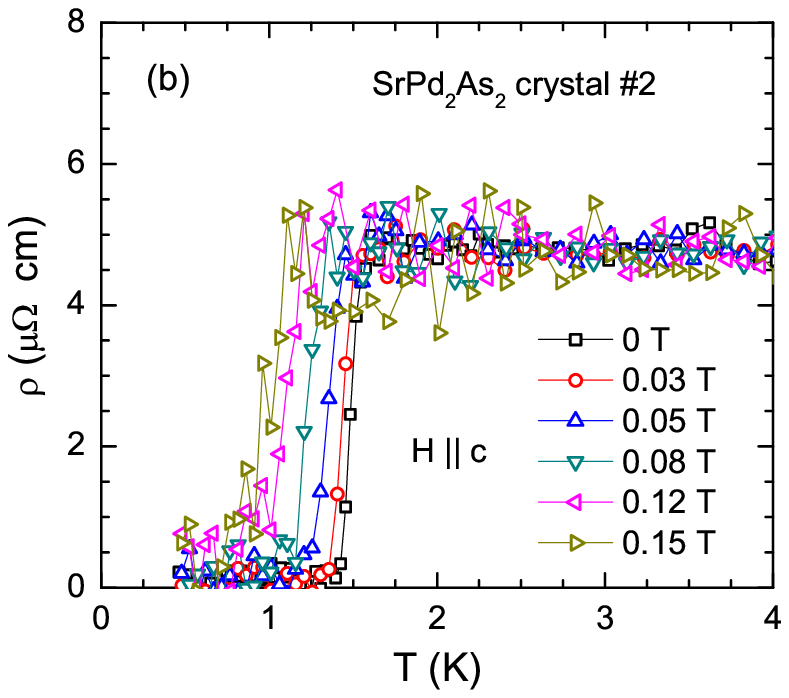}
\caption{(Color online) (a) In-plane electrical resistivity $\rho$ of a SrPd$_2$As$_2$ crystal (crystal~\#1) versus temperature $T$ measured in applied magnetic field $H=0$. The red curve is a fit by the Bloch-Gr\"{u}neisen model in Eqs.~(\ref{Eqs:BGModel}). (b) Expanded plot of the low-$T$ $\rho(T)$ data for SrPd$_2$As$_2$ crystal~\#2 for 0.45~K~$\leq T \leq$~4~K showing the superconducting transition for different $H$ applied along the $c$~axis. The noise in the data is due to the small size of the crystal and the small magnitude of $\rho$ at low temperatures.}
\label{fig:SrPd2As2_Rho}
\end{figure}

The in-plane $\rho(T)$ data for SrPd$_2$As$_2$ measured on two different crystals~\#1 and~\#2 at different $H$ are shown in Fig.~\ref{fig:SrPd2As2_Rho}.  Metallic behavior is evident from the $T$ dependence of $\rho$ in Fig.~\ref{fig:SrPd2As2_Rho}(a). The expanded plot of $\rho(T)$ in Fig.~\ref{fig:SrPd2As2_Rho}(b) reveals a superconducting transition at $T_{\rm c} = 1.5(1)$~K\@.   The $T_{\rm c}$ is suppressed with increasing $H$, as shown. The data in Fig.~\ref{fig:SrPd2As2_Rho}(b) are noisy due to the small sample size and the small voltage signal arising from the small magnitude of the resistivity at low~$T$\@.

The two crystals were found to have different residual resistivities and RRR values. For crystal \#1, $\rho_0(1.8~{\rm K}) = 7.5~\mu \Omega$\,cm and RRR  = $\rho(300\,{\rm K}/ \rho(1.8\,{\rm K}) \approx 8$, whereas for crystal \#2, $\rho_0 = 4.9~\mu \Omega$\,cm and RRR $\approx 4$. The reason the room temperature resisitivities of the two crystals are different ($\approx 60$ and~20~$\mu\Omega\,$cm, respectively) is unknown.  Since  crystal~\#1 has the higher RRR, we analyzed the $\rho(T)$ data of this crystal in Fig.~\ref{fig:SrPd2As2_Rho}(a) using the BG model. A fit of the $\rho(T)$ data in Fig.~\ref{fig:SrPd2As2_Rho}(a) by Eqs.~(\ref{Eqs:BGModel}) for 1.8~K~$\leq T \leq 300$~K gives $\rho_0 = 7.57(6)\,\mu\Omega$\,cm, $\rho(\Theta_{\rm{R}}) = 27.6(5)\,\mu\Omega$\,cm and $\Theta_{\rm{R}} = 170(3)$~K, where we used our analytic Pad\'e approximant\cite{Ryan2012} in place of the integral in Eq.~(\ref{eq:Bloch-Gruneisen}).  The good fit obtained is shown by the red curve in Fig.~\ref{fig:SrPd2As2_Rho}(a). The value of ${\cal R}$ obtained from the value of $\rho(\Theta_{\rm{R}})$ using Eq.~(\ref{eq:BG_R}) is ${\cal R} = 29.2\,\mu\Omega$\,cm. The parameters obtained from the fit are summarized in Table~\ref{Tab:RhoFitParams}.

\subsection{\label{Sec:SrPd2As2_HC}Heat Capacity}

\begin{figure}
\includegraphics[width=3.3in]{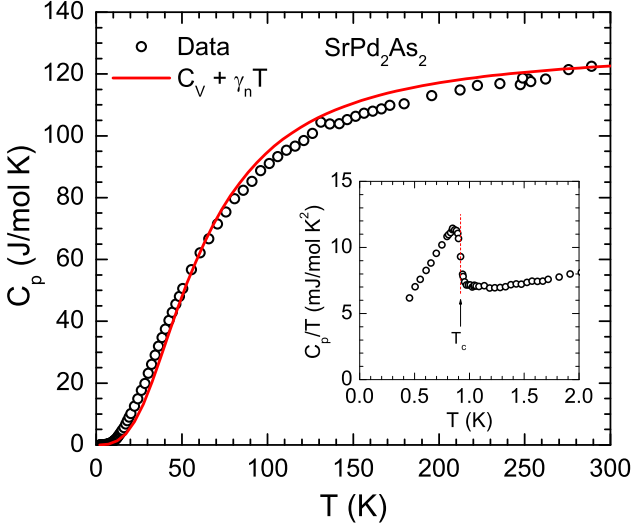}
\caption{\label{fig:SrPd2As2_HC}(Color online) The heat capacity $C_{\rm p}$ of a SrPd$_2$As$_2$ single crystal versus  temperature $T$ measured in zero magnetic field $H$\@. The red curve is the fitted sum of the contributions from the Debye lattice heat capacity $C_{\rm V\,Debye}(T)$ and predetermined electronic heat capacity $\gamma_{\rm n} T$ according to Eq.~(\ref{eq:Debye_HC-fit}). Inset: Expanded plot of $C_{\rm p}/T$ versus~$T$ for $0.45~{\rm K} \leq T \leq 2.0$~K measured on a different crystal (crystal \#2 of Fig.~\ref{fig:SrPd2As2_Rho}) at field $H = 0$. The vertical dotted red line indicates the $T_{\rm c}$.}
\end{figure}

\begin{figure}
\includegraphics[width=3.3in]{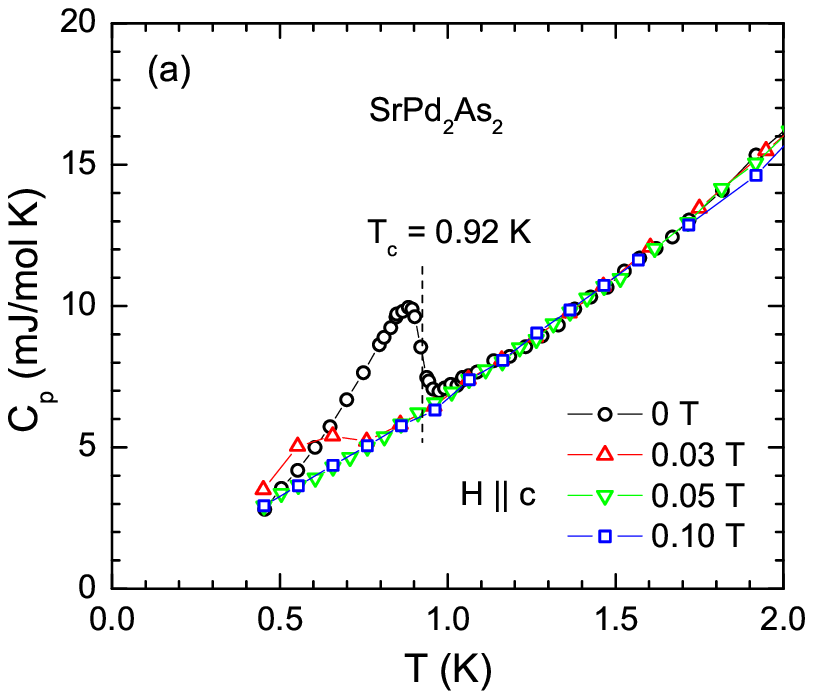}
\includegraphics[width=3.3in]{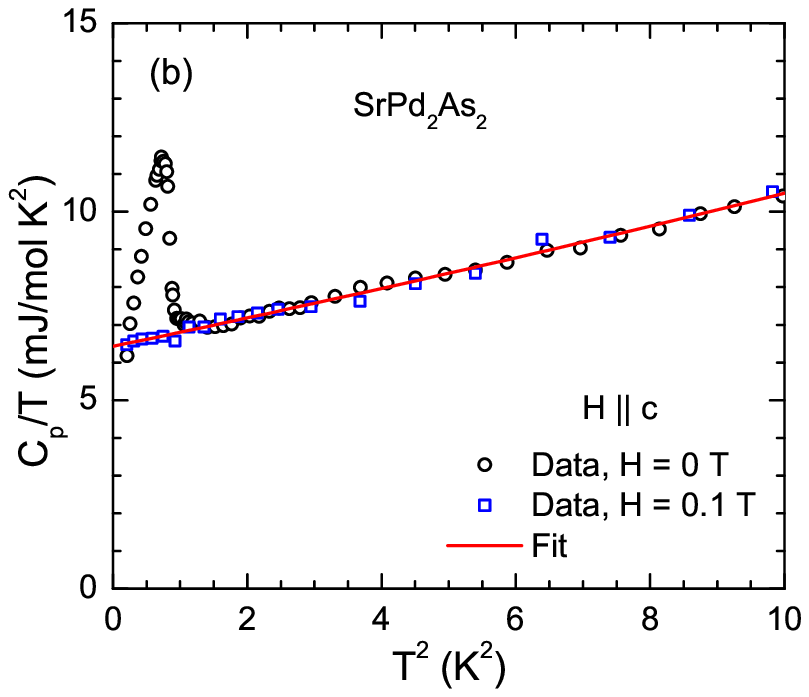}
\caption{(Color online) (a) Heat capacity $C_{\rm p}$ versus temperature $T$ of a SrPd$_2$As$_2$ single crystal [crystal \# 2 of Fig.~\ref{fig:SrPd2As2_Rho}] for 0.45~K~$\leq T \leq$~2.0~K measured at the indicated values of applied magnetic field $H$ with $H \parallel c$. (b) $C_{\rm p}/T$ vs. $T^2$ for 0.45~K~$\leq T \leq$~3.2~K with $H = 0$ and 0.1~T\@. The red curve is a fit of the $H = 0.1$~T data for $0.45~{\rm K} \leq T\leq 4.4$~K and the $H=0$ data for $1.3~{\rm K} \leq T\leq 4.4$~K by Eq.~(\ref{eq:gamma}) (the fitted data from 3.2~K to 4.4~K are not shown).}
\label{fig:SrPd2As2_HC_field}
\end{figure}

The $C_{\rm p}(T)$ data for SrPd$_2$As$_2$ are shown in Fig.~\ref{fig:SrPd2As2_HC}. The $C_{\rm p}(T= 300\,{\rm K}) \approx  124$~J/mol\,K is close to the classical Dulong-Petit high-$T$ limit $C_{\rm V} = 5R=124.7$~J/mol\,K\@. In order to correlate low-$T$ $C_{\rm p}(T)$ data with $\rho(T)$ data obtained on the same crystal, we also measured the low-$T$ $C_{\rm p}(T)$ on a different crystal~\#2 as shown in Fig.~\ref{fig:SrPd2As2_HC_field}(a), for which the $\rho(T)$ data are shown in Fig.~\ref{fig:SrPd2As2_Rho}(b). As shown in the inset of Fig.~\ref{fig:SrPd2As2_HC}, a rather sharp heat capacity jump is observed at $T_{\rm c} = 0.92(5)$~K due to the superconducting transition, where we define $T_{\rm c}$ to be the transition midpoint. Two important differences are observed between the $C_{\rm p}(T)$ and $\rho(T)$ data on crystal~\#2. First, the $T_{\rm c}$ obtained from the two measurements are different: $T_{\rm c} = 0.92(5)$~K from $C_{\rm p}(T)$ and $T_{\rm c} = 1.5(1)$~K from $\rho(T)$. Second, while the superconductivity is suppressed to a temperature below 0.45~K by $H=0.05$~T in $C_{\rm p}(T)$, superconductivity occurs at $\approx 1$~K even at $H=0.15$~T in $\rho(T)$. These two observations suggest the presence of filamentary  superconductivity in SrPd$_2$As$_2$ which is probed by $\rho(T)$ at temperatures above the bulk $T_{\rm c}$ whereas $C_{\rm p}(T)$ measures the bulk superconductivity.

A fit of the normal-state $C_{\rm p}(T)/T$ versus $T^2$ data using the $H = 0.1$~T data for $0.45~{\rm K} \leq T\leq 4.4$~K and the $H=0$ data for $1.3~{\rm K} \leq T\leq 4.4$~K in Fig.~\ref{fig:SrPd2As2_HC_field}(b) by Eq.~(\ref{eq:gamma}) (the fitted data from 3.2~K to 4.4~K are not shown) gives $\gamma_{\rm n} = 6.43(3)$~mJ/mol\,K$^2$, $\beta = 0.369(8)$~mJ/mol\,K$^4$ and $\delta = 3.7(5)~\mu$J/mol\,K$^6$ as shown by the red curve. The Debye temperature estimated from $\beta$ using Eq.~(\ref{eq:Debye-Temp}) is $\Theta_{\rm D} = 298(3)$~K\@. A fit of $C_{\rm p}(T)$ in Fig.~\ref{fig:SrPd2As2_HC} over the entire $T$ range (2--300~K) by Eqs.~(\ref{Eqs:AllTCpFit}) with $\gamma_{\rm n}$ fixed to the above value and using the Pad\'e approximant\cite{Ryan2012} in place of the Debye function gives $\Theta_{\rm D} = 245(3)$~K\@.  The fit is shown by the red curve in Fig.~\ref{fig:SrPd2As2_HC}.  The value of $\Theta_{\rm D}$ is smaller than the value of 298(3)~K obtained from the low-$T$ fit, indicating a $T$-dependent $\Theta_{\rm D}$.\cite{Ryan2012}  The parameters obtained from the analyses of the normal-state $C_{\rm p}(T)$ data are summarized in Table~\ref{tab:HCFitParams}.

The electron-phonon coupling constant is estimated from Eq.~(\ref{eq:lambda}) as $\lambda_{\rm {el-ph}} = 0.443$ using $\mu^{*} = 0.13$, $T_{\rm c} = 0.92$~K and $\Theta_{\rm D}= 298$~K\@. Then we estimate ${\cal D}_C(E_{\rm F}) = 2.73(2)$~states/(eV f.u.) for both spin directions from Eq.~(\ref{Eq:gamman}) ${\cal D}_{\rm band}(E_{\rm F}) = 1.89(1)$~states/(eV f.u.) for both spin directions from Eq.~(\ref{eq:DOS}), which is very close to that of CaPd$_2$As$_2$ in Table~\ref{tab:HCFitParams}. The Fermi velocity obtained from Eq.~(\ref{eq:vF}) is $v_{\rm F} = 1.17 \times 10^8$~cm/s, and the mean free path for this value of $v_{\rm F}$ and $\rho_{0} = 4.9~\mu\Omega$\,cm using Eq.~(\ref{eq:lvF}) is $\ell = 11.6$~nm. The $\omega_{\rm p}$ and $\lambda_{\rm L}(0)$ estimated from Eqs.~(\ref{Eq:omegapDef}) and~(\ref{Eq:lambda0}), respectively, are listed in Table~\ref{tab:SCParams2}.

\begin{figure}
\includegraphics[width=3.3in]{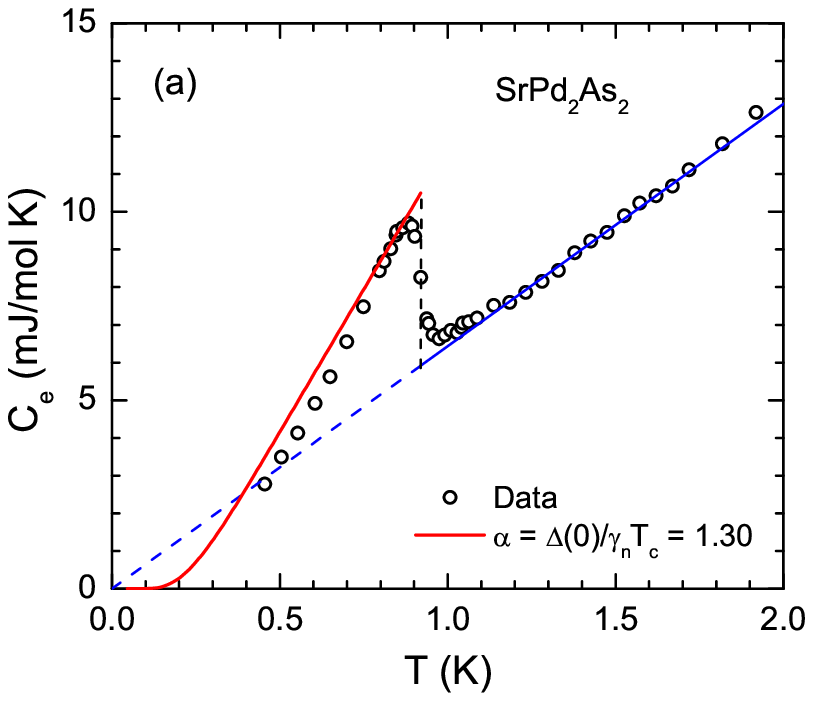}
\includegraphics[width=3.3in]{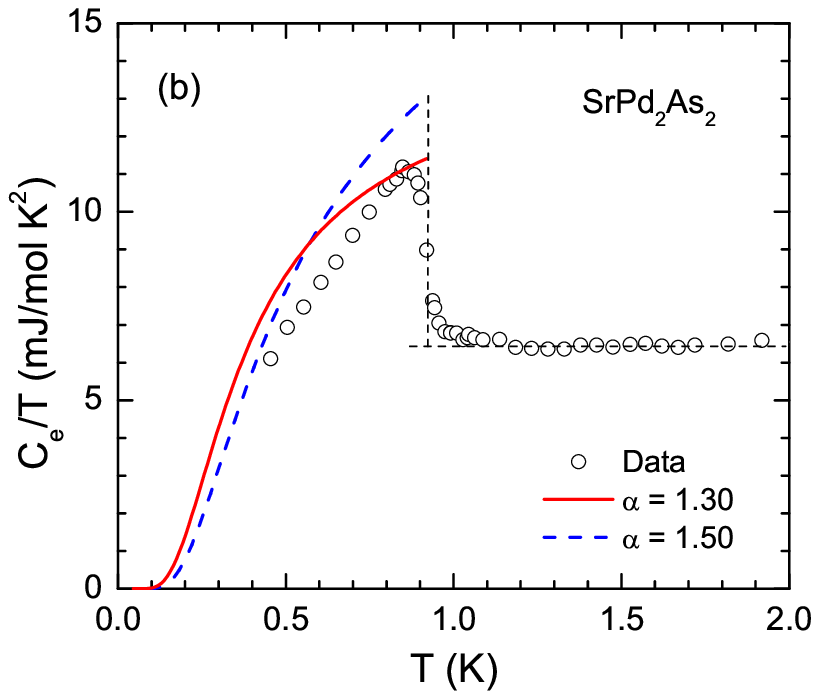}
\caption{(Color online) (a) The electronic contribution $C_{\rm e}$ versus temperature $T$ at low~$T$ obtained by subtracting the phonon contribution from the measured $C_{\rm p}(T)$ of SrPd$_2$As$_2$. (b) $C_{\rm e}/T$ versus $T$\@.  The solid red curves in (a) and~(b) are the theoretical predictions of the $\alpha$-model for $\alpha = 1.30$ and the dashed blue curve in (b) is for $\alpha = 1.50$.}
\label{fig:SrPd2As2_HC_el}
\end{figure}

\begin{figure}
\includegraphics[width=3.3in]{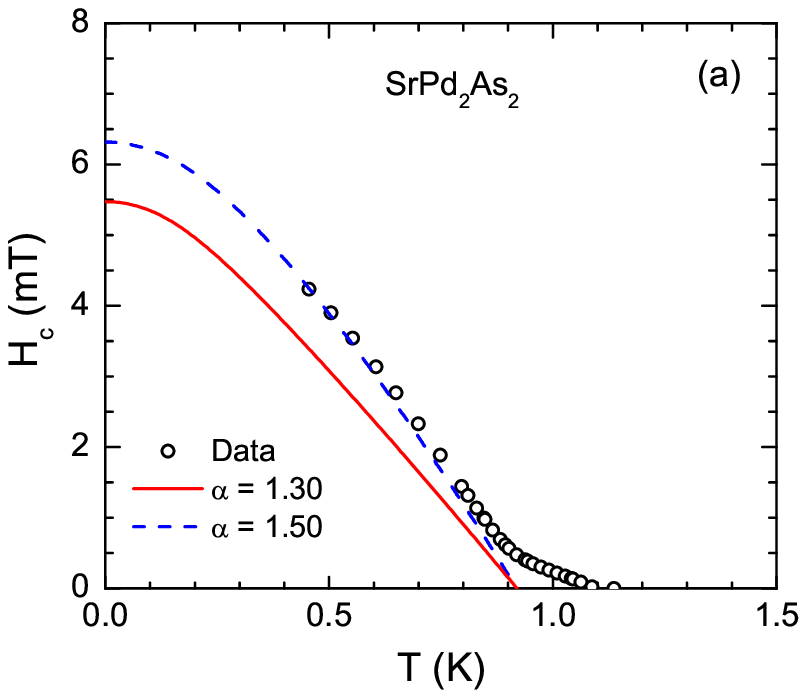}
\includegraphics[width=3.3in]{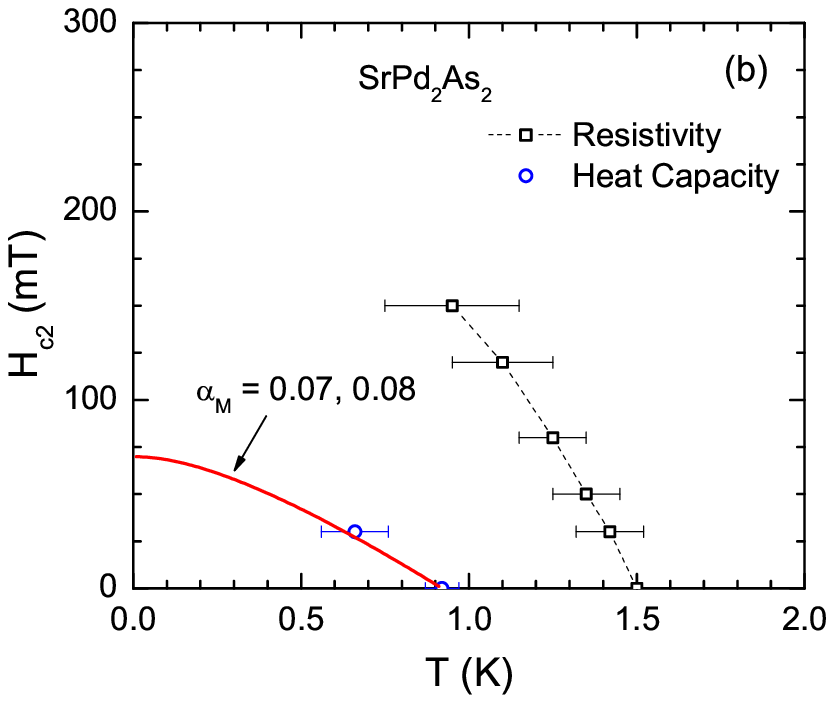}
\caption{(Color online) (a) Thermodynamic critical field $H_{\rm c}(T)$ obtained from the free energy considerations from the zero field superconducting state $C_{\rm p}(T)$ data. The solid red and dashed blue curves are the theoretical predictions of the $\alpha$-model for $\alpha = 1.30$ and~1.50, respectively. (b) Upper critical field $H_{\rm c2}(T)$ of SrPd$_2$As$_2$ determined from the electrical resistivity $\rho(T, H)$ and heat capacity $C_{\rm p}(T, H)$ data in Figs.~\ref{fig:SrPd2As2_Rho}(b) and \ref{fig:SrPd2As2_HC}(b), respectively.  The red curve is the prediction of the WHH theory in Eqs.~(\ref{Eqs:WHHEqns}) for $\alpha_{\rm M}=0.07$ and~0.08 and $\lambda_{\rm so}=0$.}
\label{fig:SrPd2As2_critical_H}
\end{figure}

The electronic contribution $C_{\rm e}(T)$ to the measured low-$T$ $C_{\rm p}(T)$ of ${\rm SrPd_2As_2}$, obtained by subtracting $\beta T^3+\delta T^5$ from $C_{\rm p}(T)$ according to Eq.~(\ref{eq:gamma}), is plotted versus~$T$ in Fig.~\ref{fig:SrPd2As2_HC_el}(a), and $C_{\rm e}(T)/T$ is plotted versus $T$ in Fig.~\ref{fig:SrPd2As2_HC_el}(b).  Utilizing the entropy-conserving construction in Fig.~\ref{fig:SrPd2As2_HC_el}(b) we obtain $\Delta C_{\rm e}(T_{\rm c})/T_{\rm c} = 5.0(2)$~mJ/mol\,K$^2$, $\Delta C_{\rm e}(T_{\rm c}) = 4.6(2)$~mJ/mol\,K using $T_{\rm c} = 0.92(5)$~K and $\Delta C_{\rm e}(T_{\rm c})/ \gamma_{\rm n} T_{\rm c} = 0.77(5)$ using $\gamma_{\rm n} =6.43(3)$~mJ/mol\,K$^2$.  The value of $\Delta C_{\rm e}(T_{\rm c})/ \gamma_{\rm n} T_{\rm c}$ is significantly smaller than the BCS weak-coupling value of 1.43, as was also the case for CaPd$_2$As$_2$ discussed above, and from Eq.~(\ref{Eq:DelCgamTcDelta}) we obtain $\alpha = 1.30(4)$ which may be compared with the BCS value of 1.764. In Figs.~\ref{fig:SrPd2As2_HC_el}(a) and~\ref{fig:SrPd2As2_HC_el}(b) we show as the red curves the respective theoretical predictions of the $\alpha$-model obtained using $\alpha=1.30$ in Eqs.~(\ref{Eqs:CesCalcs}).  Also shown as the dashed blue curve in Fig.~\ref{fig:SrPd2As2_HC_el}(b) is the theoretical prediction for $\alpha=1.50$, which is the value that best fits the $H_{\rm c}(T)$ data in Fig.~\ref{fig:SrPd2As2_critical_H}(a) below.

The $H_{\rm c}(T)$ for ${\rm SrPd_2As_2}$ is obtained by integrating the $C_{\rm e}(T)$ data in Fig.~\ref{fig:SrPd2As2_HC_el} according to Eqs.~(\ref{Eqs:HcFromSe}) and the results are shown in Fig.~\ref{fig:SrPd2As2_critical_H}(a).  The shoulder just above $T_{\rm c}$ arises from the high-$T$ shoulder in the $C_{\rm e}(T)$ data in Figs.~\ref{fig:SrPd2As2_HC_el}(a) and~\ref{fig:SrPd2As2_HC_el}(b).  The value of $H_{\rm c}(0)$ is calculated from Eq.~(\ref{Eq:HcFromGammaTc}) using $\alpha = 1.30$, $\gamma_{\rm n} = 6.43$~mJ/mol\,K$^2$ and $T_{\rm c} = 0.92$~K, yielding $H_{\rm c}(0)= 5.47$~mT\@. The theoretical prediction of the $\alpha$-model in Eqs.~(\ref{Eqs:Hc(T)Calc}) for $H_{\rm c}(T)$ using these parameters is plotted in Fig.~\ref{fig:SrPd2As2_critical_H}(a). Although the $T$ dependence of the data is reproduced, the calculated magnitude does not agree with the data.  A better fit as shown in Fig.~\ref{fig:SrPd2As2_critical_H}(a) by the blue dashed curve is obtained using $\alpha = 1.50$, for which $H_{\rm c}(0)$ is calculated as above to be 6.3~mT\@.  

The $H_{\rm c2}$ versus~$T$ data obtained from the above $C_{\rm p}(T)$ and $\rho(T)$ measurements with $H \parallel c$ are shown in Fig.~\ref{fig:SrPd2As2_critical_H}(b).  As already mentioned the two measurements show different $T_{\rm c}$'s in zero field.  Consistent with this difference, the $H_{\rm c2}(T)$ behavior derived from the $\rho(T)$ data indicates that the upper critical field of the filamentary superconductivity is larger than the bulk $H_{\rm c2}$.  The above values of $H_{\rm c}(0)$ for both $\alpha=1.30$ or~1.50 are much smaller than the bulk $H_{\rm c2}(0)$ extrapolated from the heat capacity data in Fig.~\ref{fig:SrPd2As2_critical_H}(b), indicating type-II superconductivity in SrPd$_2$As$_2$ as was also found above to be the case in CaPd$_2$As$_2$.

From the $C_{\rm p}(T)$ data in Fig.~\ref{fig:SrPd2As2_critical_H}(b) one obtains $dH_{\rm c2}(T)/dT|_{T=T_{\rm c}} = - 0.12(4)$~T/K\@.  Equation~(\ref{Eq:Hc2Orb}) then gives $H_{\rm c2}^{\rm Orb}(0) = 0.077$~T in the clean limit and $H_{\rm c2}^{\rm Orb}(0) = 0.073$~T in the dirty limit. Using $\alpha=1.30$, the Pauli-limiting field is obtained from Eq.~(\ref{Hp(0)}) as $H_{\rm P}(0) = 1.37$~T and the Maki parameter in Eq.~(\ref{Eq:alphaMDef}) is then $\alpha_{\rm M} = 0.08$. An estimate of $H_{\rm c2}(T)$ obtained from the WHH prediction in Eqs.~(\ref{Eqs:WHHEqns}) using $\alpha_{\rm M} = 0.08$ and~$\lambda_{\rm so}=0$ is shown as the solid red curve in Fig.~\ref{fig:SrPd2As2_critical_H}(b), from which we obtain $H_{\rm c2}(0) = 70$~mT\@.

The Ginzburg-Landau parameter is obtained from Eq.~(\ref{Eq:KappaGL}) using $H_{\rm c2}(0)=70$~mT and $H_{\rm c}(0)=5.5$~mT, yielding $\kappa_{\rm GL} = 9.0$. The lower critical field estimated from Eq.~(\ref{Eq:Hc1fromKappaGL}) is $H_{\rm c1}(0) = 0.94$~mT\@. The Ginzburg-Landau coherence length at $T=0$ obtained from Eq.~(\ref{Eq:xiFromHc2}) is $\xi(0) = 69$~nm which together with $\ell =11.6$~nm and Eq.~(\ref{eq:xi0l}) for the dirty limit gives $\xi_{0} = 509$~nm. The Fermi velocity estimated from $\xi_{0}$ using Eq.~(\ref{eq:xivF}) is $v_{\rm F} = 0.25 \times 10^8$~cm/s for $\alpha = 1.30$ which is of the same order as the above estimated value of $v_{\rm F}$ from the density of states. Corresponding values of the above parameters for $\alpha=1.50$ were also calculated.  A summary of the measured and derived superconducting parameters for SrPd$_2$As$_2$ is given in Table~\ref{tab:SCParams2}.

\begin{table}
\caption{\label{tab:SCParams2} Measured and derived superconducting and relevant normal state parameters for SrPd$_2$As$_2$.  $T_{\rm c}$: bulk superconducting transition temperature; $\gamma_{\rm n}$: observed Sommerfeld coefficient of the linear term in the low-$T$ normal-state heat capacity; $\lambda_{\rm el-ph}$: electron-phonon coupling constant; $\ell$: mean free path at low~$T$; $\omega_{\rm p}$: plasma angular frequency; $\lambda_{\rm L}$: London penetration depth; $\lambda_{\rm eff}$: magnetic penetration depth; $H_{\rm c2}^{\rm Orb}$: orbital upper critical magnetic field; $\Delta$: superconducting order parameter; $\alpha = \Delta(0)/k_{\rm B}T_{\rm c}$; $\Delta C_{\rm e}$: heat capacity jump at $T_{\rm c}$; $H_{\rm P}$: Pauli limiting upper critical field; $\alpha_{\rm M}$: Maki parameter; $H_{\rm c}$, $H_{\rm c1}$,  $H_{\rm c2}$: thermodynamic, lower critical, and fitted upper critical magnetic fields, respectively; $\kappa_{\rm GL}$: Ginzburg-Landau parameter; $\xi$: Ginzburg-Landau coherence length; $\xi_0$: BCS superconducting coherence length.  }
\begin{ruledtabular}
\begin{tabular}{lcc}
SrPd$_2$As$_2$ property & \multicolumn{2}{c}{value} \\
\hline
$T_{\rm c}$ (K)                  & \multicolumn{2}{c}{0.92(5)}  \\
$\gamma_{\rm n}$ (mJ/mol\,K$^{2}$)  & \multicolumn{2}{c}{6.43(3)}  \\
$\lambda_{\rm el-ph}$             & \multicolumn{2}{c}{0.443}  \\
$\ell~(m^\ast=m_{\rm e})$ (nm)   & \multicolumn{2}{c}{11.6}  \\
$\omega_{\rm p}~(m^\ast=m_{\rm e})~(10^{16}~{\rm rad/s})$  & \multicolumn{2}{c}{1.52}\\
$\lambda_{\rm L}^{\rm calc}(0)$~(clean limit) (nm)         & \multicolumn{2}{c}{19.7}\\
$\lambda_{\rm eff}^{\rm obs}(0)$ (nm)                        & \multicolumn{2}{c}{170(70)} \\
$H_{\rm c2}^{\rm Orb}(T=0)$~(dirty limit) (T)  & \multicolumn{2}{c}{0.073}\\
$\Delta(0)/k_{\rm B}$ (K) (observed)	&		\multicolumn{2}{c}{2.05(20)}	\\

                                 & $\alpha =1.30$  & $\alpha=1.50$ \\
\hline
$\Delta C_{\rm e}$ (mJ/mol\,K)   &  4.6(2) & 6.1(4)\\
$\Delta C_{\rm e}/\gamma_{\rm n} T_{\rm c}$  & 0.77(5) & 1.03(8)\\
$H_{\rm P} (0)$ (T)              & 1.37     & 1.58 \\
$\alpha_{\rm M}$                 & 0.08     & 0.07 \\
$H_{\rm c}(T=0)$ (mT)            & 5.5     & 6.3 \\
$H_{\rm c1}(T=0)$ (mT)           & 0.94     & 1.17\\
$H_{\rm c2}(T=0)$ (mT)           & 70     & 70\\
$\kappa_{\rm GL}$                & 9.0     & 7.8 \\
$\xi(T=0)$ (nm)                  & 69     & 69 \\
$\xi_0$ (nm)                     & 509      & 509 \\
$\lambda_{\rm eff}^{\rm calc}(0)$~(dirty limit) (nm)       & 130 & 130\\
\end{tabular}
\end{ruledtabular}
\end{table}

\subsection{\label{Sec:SrPd2As2_M(H,T)} Magnetization and Magnetic Susceptibility}
\begin{figure}
\includegraphics[width=3.3in]{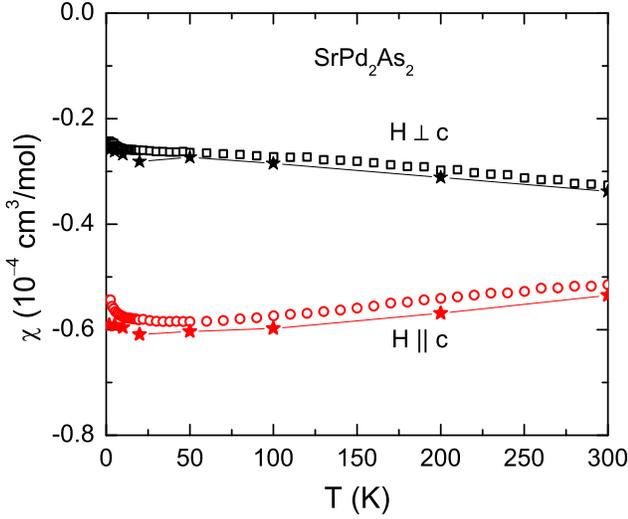}
\caption{(Color online) Zero-field-cooled magnetic susceptibility $\chi$ of a ${\rm SrPd_2As_2}$ single crystal versus temperature $T$ in a magnetic field $H = 3.0$~T applied along the $c$-axis ($\chi_c,\ H \parallel c$) and in the $ab$-plane ($\chi_{ab},\ H \perp c$). The filled stars represent the intrinsic $\chi$ obtained in the Appendix from fitting $M(H)$ isotherm data by Eq.~(\ref{eq:MH_linear-fit}). The lines joining the stars are guides to the eye.}
\label{fig:MT_SrPd2As2}
\end{figure}

The ZFC $\chi(T)\equiv M(T)/H$ data for a SrPd$_2$As$_2$ single crystal measured in $H$ = 3.0~T are shown in Fig.~\ref{fig:MT_SrPd2As2} together with the intrinsic $\chi$ obtained from fitting $M(H)$ isotherm data for $H \geq 2$~T in the Appendix by Eq.~(\ref{eq:MH_linear-fit}). These two data sets are in excellent agreement over the whole $T$~range, indicating the near absence of ferromagnetic and saturable paramagnetic impurities in the crystal.  The very small upturns below $\sim 10$~K in $\chi(T)$ in Fig.~\ref{fig:MT_SrPd2As2} are attributed to a trace amount of paramagnetic impurities.  The $\chi$ is diamagnetic and exhibits a weak $T$-dependence with a strong anisotropy $\chi_{ab} > \chi_{c}$. The large anisotropy in $\chi$ most likely originates from anisotropy in the paramagnetic Van Vleck orbital contribution $\chi_{\rm VV}$ in Eq.~(\ref{eq:chi}), although the strong temperature dependence of the anisotropy is unusual and unexpected for such a compound and for which the origin is not clear.  Despite having the same crystal structure, the $\chi$ anisotropy in SrPd$_2$As$_2$ is opposite to that of CaPd$_2$As$_2$ in Fig.~\ref{fig:MT_CaPd2As2} for which we observed $\chi_{ab} < \chi_{c}$.  This difference in the sign of the anisotropy between the two compounds is evidently again attributable to a difference in the sign of the anisotropy in the Van Vleck contributions between the two compounds.

The powder and temperature (1.8--300~K) average of the intrinsic $\chi$ obtained from the $M(H)$ isotherms in the Appendix is $\langle \chi\rangle = -3.9 \times 10^{-5}$~cm$^3$/mol. The diamagnetic core susceptibility estimated using the atomic diamagnetic susceptibilities \cite{Mendelsohn1970} is $\chi_{\rm {core}} = -1.93 \times 10^{-4}$~cm$^3$/mol. The Pauli susceptibility estimated from Eq.~(\ref{eq:Chi-Pauli}) using ${\cal D}_{\rm band}(E_{\rm F})$ = 1.89~states/eV\,f.u.\ for both spin directions is $\chi_{\rm {P}} = 6.1 \times 10^{-5}$~cm$^3$/mol.  The $\chi_{\rm {L}} = -2.0 \times 10^{-5}$~cm$^3$/mol was obtained by taking $m_{\rm band}^* =  m_{\rm e}$ in Eq.~(\ref{eq:Chi-Landau}). Then $\langle\chi_{\rm VV}\rangle$ is obtained from these values using Eq.~(\ref{eq:chi}).  The various $\chi$ contributions are summarized in Table~\ref{tab:Chi_Contr}.

\section{\label{Sec:LondonPD} Magnetic Penetration Depth Measurements of C\lowercase{a}P\lowercase{d}$_2$A\lowercase{s}$_2$ and S\lowercase{r}P\lowercase{d}$_2$A\lowercase{s}$_2$}

\begin{figure}[t]
\includegraphics[width=3.3in]{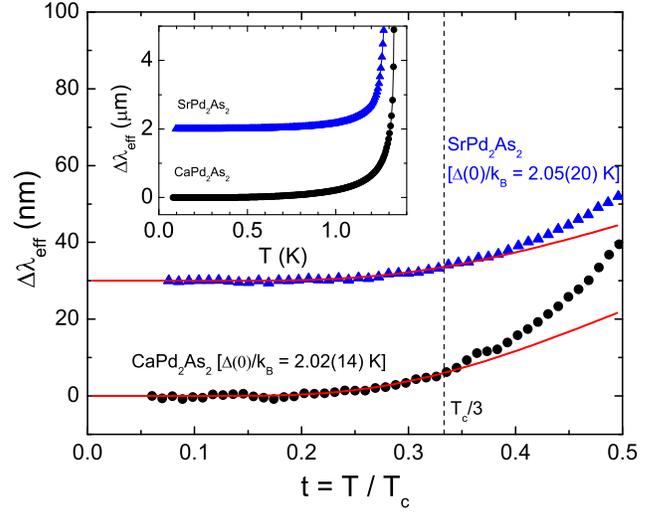}
\caption{The change $\Delta\lambda_{\rm eff} \equiv \lambda_{\rm eff}(T)-\lambda_{\rm eff}(T\to0)$ of the magnetic penetration depth $\lambda_{\rm eff}$ measured in CaPd$_2$As$_2$ (filled circles) and SrPd$_2$As$_2$ (filled triangles). The data for SrPd$_2$As$_2$ are shifted vertically upwards by 30~nm for clarity. The solid curve for each compound is the best fit of the data by the prediction in Eq.~(\ref{Eq:DeltaLambda}) for a single-gap $s$-wave BCS superconductor for $T\lesssim T_{\rm c}/3$.   The vertical dashed line is the upper temperature limit for the fits.  Inset: $\Delta\lambda_{\rm eff}(T)$ for both compounds up to $T=T_{\rm c}$.  The data for SrPd$_2$As$_2$ are all shifted upwards by $2~\mu$m for clarity.}
\label{fig:lambda}
\end{figure}

Figure~\ref{fig:lambda} shows the temperature variation of the $ab$-plane magnetic penetration depth, $\Delta\lambda_{\rm eff}(T)\equiv \lambda_{\rm eff}(T)-\lambda_{\rm eff}(0)$, measured in CaPd$_2$As$_2$ and SrPd$_2$As$_2$ crystals, represented by open circles and triangles, respectively. The absolute value of the penetration depth was obtained using the TDR technique by matching the frequency shift, $\Delta f(T)$, to the skin depth, $\delta$, calculated from the resistivity. The superconducting transition temperature was determined as the temperature of the maximum of $d\Delta\lambda_{\rm eff}/dT$. The determined $T_{\rm c}$'s are 1.34~K and 1.26~K for CaPd$_2$As$_2$ and SrPd$_2$As$_2$, respectively.  These values are higher than the bulk $T_{\rm c}$'s of 1.27(3)~K and 0.92(5)~K determined from respective $C_{\rm p}(T)$ data (Tables~\ref{tab:SCParams} and \ref{tab:SCParams2}, respectively). Even so, the actual onset of the diamagnetic response is observed at even higher temperatures, $T_{\rm c}^{\rm onset}=1.50$~K and $1.72$~K for CaPd$_2$As$_2$ and SrPd$_2$As$_2$, respectively. The $T$ dependences of $\Delta\lambda_{\rm eff}$ for the two compounds up to $T_{\rm c}$ are shown in the inset of Fig.~\ref{fig:lambda}.

At low temperatures, the $\Delta\lambda_{\rm eff}(T)$ in Fig.~\ref{fig:lambda} of each sample shows a clear saturation on cooling, which is an indication of a fully-gapped superconducting order parameter in both compounds. Our compounds are dirty-limit superconductors, for which the magnetic penetration depth in the single-band model for $T/T_{\rm c}\ll 1$ in Tinkham's notation\cite{Tinkham1996} is 
\bse
\be
\lambda_{\rm eff}(T) = \lambda_{\rm L}(T)\sqrt{1+\frac{\xi_0}{\ell}},
\ee
where $\lambda_{\rm L}(T)$ is the clean-limit BCS London penetration depth prediction, yielding the $T=0$ expression in Eq.~(\ref{eq:lambda_eff}), and also for $T/T_{\rm c}\ll 1$ the  expression
\be
\frac{\lambda_{\rm eff}(T)}{\lambda_{\rm eff}(0)} = \frac{\lambda_{\rm L}(T)}{\lambda_{\rm L}(0)}.
\ee
Defining $\Delta \lambda(T) = \lambda(T) - \lambda(0)$, one obtains
\be
\frac{\Delta\lambda_{\rm eff}(T)}{\lambda_{\rm eff}(0)} = \frac{\Delta\lambda_{\rm L}(T)}{\lambda_{\rm L}(0)}.
\label{Eq:lefflL}
\ee
\ese
The right-hand side of this equation is just the clean-limit BCS prediction for local electrodynamics given by\cite{Prozorov2006, Bardeen1957,Johnston2013}
\begin{equation}
\frac{\Delta\lambda_{\rm L}(T)}{\lambda_{\rm L}(0)}=\sqrt{\frac{\pi\Delta(0)}{2k_{\rm B}T}}\ \exp\left[-\frac{\Delta(0)}{k_{\rm B}T}\right].
\label{Eq:DeltaLambda}
\end{equation}
Combining Eqs.~(\ref{Eq:lefflL}) and~(\ref{Eq:DeltaLambda}) gives
\begin{equation}
\Delta\lambda_{\rm eff}(T)=\lambda_{\rm eff}(0)\sqrt{\frac{\pi\Delta(0)}{2k_{\rm B}T}}\ \exp\left[-\frac{\Delta(0)}{k_{\rm B}T}\right].
\label{Eq:DeltaLambda2}
\end{equation}

The experimental data are fitted well up to $T\approx T_{\rm c}/3$ by Eq.~(\ref{Eq:DeltaLambda2}) as shown by the solid curves in Fig.~\ref{fig:lambda}, where the fitting parameters are $\lambda_{\rm eff}(0) = 210\pm60$~nm and $\Delta(0)/k_{\rm B} = 2.02\pm0.14$~K for CaPd$_2$As$_2$ and $\lambda_{\rm eff}(0) = 170 \pm 70$~nm and $\Delta(0)/k_{\rm B} = 2.05\pm0.20$~K for SrPd$_2$As$_2$.  The listed errors are systematic errors obtained from the spread of the fitting parameters depending on different choices of the upper temperature limit near $T_{\rm c}/3$.

Using the bulk $T_{\rm c}$ values in Tables~\ref{tab:SCParams} and~\ref{tab:SCParams2} and the above values of $\Delta(0)/k_{\rm B}$, we obtain $\alpha = 1.59(14)$ for CaPd$_2$As$_2$ and $\alpha = 2.23(0.32)$ for SrPd$_2$As$_2$. The value of $\alpha$ for CaPd$_2$As$_2$ is identical within the error bars to the value of 1.58(2) in Eq.~(\ref{Eq:alphaCaPd2As2}) that was determined from the heat capacity jump, both of which are smaller than the value $\alpha_{\rm BCS}\approx 1.764$ expected for an isotropic weak-coupling BCS superconductor.\cite{Bardeen1957,Johnston2013}  This reduction is most likely due to a moderate anisotropy of the order parameter\cite{Johnston2013} rather than multiple order parameters, because well-known multi-gap superconductors such as MgB$_2$, \cite{Fletcher2005} NbSe$_2$, \cite{Fletcher2007} and LiFeAs (Ref.~\onlinecite{Kim2011}) have shown much lower values of $\alpha$ for the smaller gap.  The accuracy of $\alpha$ for SrPd$_2$As$_2$ is uncertain because of the significantly larger superconducting transition width obtained from the heat capacity measurements for this compound, and will therefore not be further considered.

The above parameter values obtained for CaPd$_2$As$_2$ and SrPd$_2$As$_2$ are listed in Tables~\ref{tab:SCParams} and~\ref{tab:SCParams2}, respectively.

\section{\label{BaPd2As2} Physical Properties of B\lowercase{a}P\lowercase{d}$_2$A\lowercase{s}$_2$ Crystals}

\subsection{\label{Sec:BaPd2As2_Rho} Electrical Resistivity}

\begin{figure}
\includegraphics[width=3.3in]{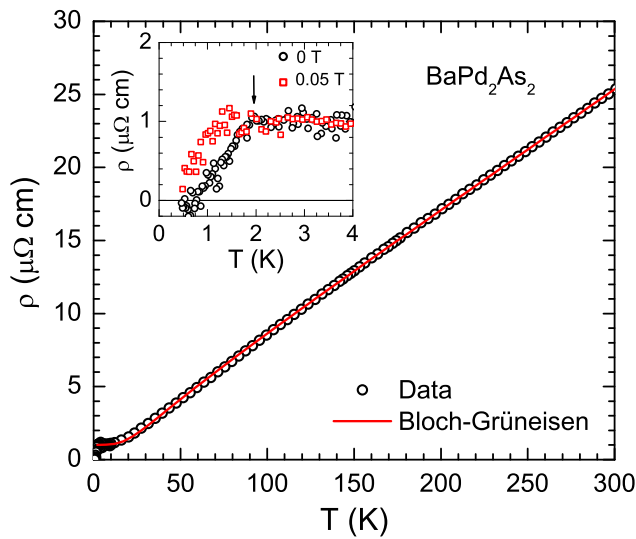}
\caption{(Color online) In-plane electrical resistivity $\rho$ of a BaPd$_2$As$_2$ crystal versus temperature $T$  measured in zero magnetic field~$H$\@. The red curve is a fit by the Bloch-Gr\"{u}neisen model. Inset: Expanded plot of the low-$T$ $\rho(T)$ data showing the onset of superconductivity at $T\lesssim 2.0$~K measured at $H = 0$ and 0.05~T applied along the $c$-axis.  We infer that this superconductivity is filamentary and not bulk.}
\label{fig:BaPd2As2_Rho}
\end{figure}

The in-plane $\rho(T)$ data for a BaPd$_2$As$_2$ crystal are shown in Fig.~\ref{fig:BaPd2As2_Rho}. The $T$ dependence of $\rho$ reveals metallic behavior with a very small residual resistivity $\rho_0 \approx 1~\mu \Omega$\,cm and a large RRR $\approx 25$, demonstrating the high quality of the crystal.  The scale of the resistivity for BaPd$_2$As$_2$ is smaller and the RRR is much larger than those of CaPd$_2$As$_2$ and SrPd$_2$As$_2$.  The $\rho(T)$ data were fitted by Eqs.~(\ref{Eqs:BGModel}) for 2~K~$\leq T \leq$~300~K using the analytic Pad\'e approximant function\cite{Ryan2012} as shown by red curve in Fig.~\ref{fig:BaPd2As2_Rho}. The excellent fit obtained yielded the fitting parameters $\rho_0 = 1.02(1)\,\mu\Omega$\,cm, $\rho(\Theta_{\rm{R}}) = 8.84(1)\,\mu\Omega$\,cm and $\Theta_{\rm{R}} = 114(1)$~K\@. The value of the constant $\mathcal{R}$ obtained using Eq.~(\ref{eq:BG_R}) is $\mathcal{R} = 9.34\,\mu\Omega$\,cm. The fit parameters are summarized in Table~\ref{Tab:RhoFitParams}.

The expanded low-$T$ plot of $\rho(T)$ in the inset of Fig.~\ref{fig:BaPd2As2_Rho} reveals an onset of superconductivity at $T_{\rm c\,  onset} \approx 2.0$~K for $H=0$ with zero resistance at about 0.6~K.  However, the transition width is very large compared to those of CaPd$_2$As$_2$ and SrPd$_2$As$_2$, and the heat capacity measurements in the following section show no evidence for superconductivity above 0.45~K\@. An applied field of 0.05~T decreases $T_{\rm c\,  onset}$ by $\approx 0.8$~K, as shown.

\subsection{\label{Sec:BaPd2As2_HC}Heat Capacity}

\begin{figure}
\includegraphics[width=3.3in]{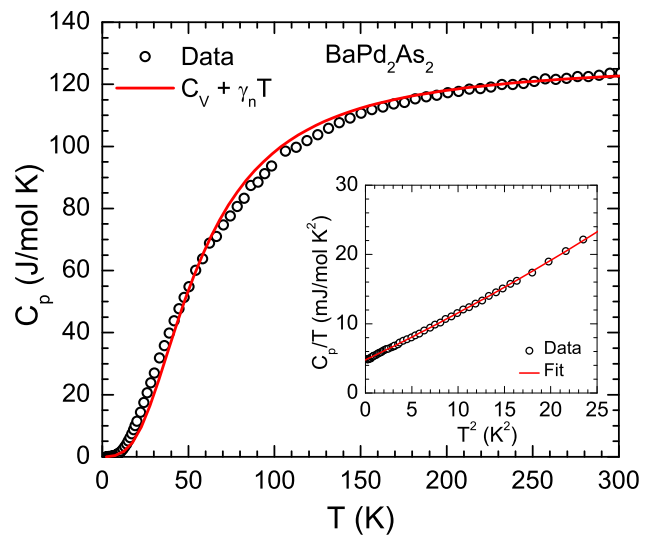}
\caption{(Color online) Heat capacity $C_{\rm p}$ of a BaPd$_2$As$_2$ single crystal versus temperature $T$ for 2.4~K~$\leq T \leq$~300~K measured in zero magnetic field. The red solid curve is the fitted sum of the contributions from the Debye lattice heat capacity $C_{\rm V\,Debye}(T)$ and predetermined electronic heat capacity $\gamma_{\rm n} T$ according to Eq.~(\ref{eq:Debye_HC-fit}). Inset: $C_{\rm p}/T$ versus $T^2$ for $0.45~{\rm K} \leq T \leq 5$~K\@. The red curve is a fit of the data by Eq.~(\ref{eq:gamma}) for $0.45~{\rm K} \leq T\leq 5$~K.}
\label{fig:BaPd2As2_HC}
\end{figure}

The $C_{\rm p}(T)$ data for BaPd$_2$As$_2$ are shown in Fig.~\ref{fig:BaPd2As2_HC}. Like the other two compounds, the $C_{\rm p}(T= 300~{\rm K}) = 124$~J/mol\,K is close to the expected high-$T$ classical Dulong-Petit value.  An expanded low-$T$ plot of $C_{\rm p}/T$ versus $T^2$ is shown in the inset of Fig.~\ref{fig:BaPd2As2_HC}.  A fit of the data with $0.45~{\rm K} \leq T\leq 5$~K by Eq.~(\ref{eq:gamma}) yields $\gamma_{\rm n} = 4.79(2)$~mJ/mol\,K$^2$, $\beta = 0.638(5)$~mJ/mol\,K$^4$ and $\delta = 4.0(3)~\mu$J/mol\,K$^6$, as shown by the red curve in the inset. A ${\cal D}_C(E_{\rm F}) = 2.03(1)$~states/(eV f.u.) for both spin directions is estimated from $\gamma_{\rm n}$ using Eq.~(\ref{Eq:gamman}). The value $\Theta_{\rm D} = 248(1)$~K is obtained from $\beta$ using Eq.~(\ref{eq:Debye-Temp}).

A value $\Theta_{\rm D} = 227(2)$~K is obtained by fitting the $C_{\rm p}(T)$ data by Eqs.~(\ref{Eqs:AllTCpFit}) over the entire $T$ range (2--300~K), as shown by the red curve in Fig.~\ref{fig:BaPd2As2_HC}.  Here again we used analytic Pad\'{e} approximant function \cite{Ryan2012} for $C_{\rm V\,Debye}(T)$ and set $\gamma_{\rm n}$ to the fixed value obtained above from the fit to the low-$T$ $C_{\rm p}(T)$ data. The normal-state parameters obtained from these fits are summarized in Table~\ref{tab:HCFitParams}.

In contrast to the observation of the onset of superconductivity in the $\rho(T)$ data at $\approx 2.0$~K in the inset of Fig.~\ref{fig:BaPd2As2_Rho}, no corresponding feature is observed in the bulk $C_{\rm p}(T)$ data above 0.45~K in the inset of Fig.~\ref{fig:BaPd2As2_HC}, which indicates that there is no bulk superconductivity in BaPd$_2$As$_2$ and hence the superconductivity detected by the $\rho(T)$ measurements is filamentary in nature.

\subsection{\label{Sec:BaPd2As2_M(H,T)} Magnetization and Magnetic Susceptibility}

\begin{figure}
\includegraphics[width=3.3in]{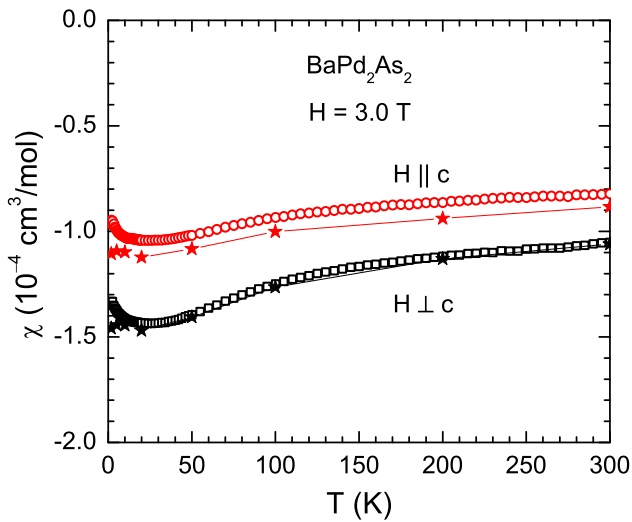}
\caption{(Color online) Zero-field-cooled magnetic susceptibility $\chi$ of a ${\rm BaPd_2As_2}$ single crystal as a function of temperature $T$ in the temperature range 1.8--300~K measured in a magnetic field $H = 3.0$~T applied along the $c$~axis ($\chi_c,\ H \parallel c$) and in the $ab$~plane ($\chi_{ab},\ H \perp  c$). The filled stars are the intrinsic $\chi$ obtained in the Appendix from fitting $M(H)$ isotherm data by Eq.~(\ref{eq:MH_linear-fit}). The lines joining the stars are guides to the eye.}
\label{fig:MT_BaPd2As2}
\end{figure}

The anisotropic $\chi(T)\equiv M(T)/H$ data of a ${\rm BaPd_2As_2}$ single crystal in $H$ = 3.0~T are shown in Fig.~\ref{fig:MT_BaPd2As2}. The $\chi$ is diamagnetic over the whole $T$ range and exhibits a weak $T$-dependence. We find that $\chi_c > \chi_{ab}$, which has the same sign of the $\chi$ anisotropy as in ${\rm CaPd_2As_2}$ but is opposite to that of ${\rm SrPd_2As_2}$.  One must keep in mind in making these comparisons that the crystal structure of ${\rm BaPd_2As_2}$ is different from that of ${\rm CaPd_2As_2}$ and ${\rm SrPd_2As_2}$.  The intrinsic anisotropic susceptibilities obtained from the slopes of high-field linear fits to the $M(H)$ isotherms in the Appendix are shown by solid stars in Fig.~\ref{fig:MT_BaPd2As2}.  These data are in rather good agreement with the $\chi(T)\equiv M(T)/H$ data in Fig.~\ref{fig:MT_BaPd2As2} above 25~K, and indicate that the low-$T$ upturns below 25~K are due to a small amount of paramagnetic impurities.

The different contributions to the intrinsic $\chi$ of ${\rm BaPd_2As_2}$ were estimated following the same approach as for the Ca and Sr members above. The powder- and temperature-average of the intrinsic $\chi$ obtained from the analyses of the $M(H)$ isotherms from 1.8 to 300~K in the Appendix is $\langle \chi\rangle = -1.24 \times 10^{-4}$~cm$^3$/mol. We also obtain $\chi_{\rm {core}} = -2.16 \times 10^{-4}$~cm$^3$/mol, $\chi_{\rm {P}} = 6.6 \times 10^{-5}$~cm$^3$/mol assuming ${\cal D}_{\rm band}(E_{\rm F}) = {\cal D}_C(E_{\rm F})$, and $\chi_{\rm {L}} = -2.2 \times 10^{-5}$~cm$^3$/mol [using $m_{\rm band}^* =  m_{\rm e}$ in Eq.~(\ref{eq:Chi-Landau})]. A value $\langle\chi_{\rm {VV}}\rangle = 4.8 \times 10^{-5}$~cm$^3$/mol is then obtained using Eq.~(\ref{eq:chi}). These contributions are summarized in Table~\ref{tab:Chi_Contr}.

\section{\label{Conclusion} Summary and Conclusions}

The crystallographic, electronic transport, thermal, magnetic and superconducting properties of $A$Pd$_2$As$_2$ ($A$ = Ca, Sr, Ba) single crystals were investigated. The magnetic measurements in the normal state reveal anisotropic diamagnetism with $\chi_{c} > \chi_{ab}$ for CaPd$_2$As$_2$ and BaPd$_2$As$_2$, and $\chi_{ab} > \chi_{c}$ for SrPd$_2$As$_2$.  The $\chi(T)$, $\rho (T)$ and $C_{\rm p}(T)$ data indicate $sp$-band-like metallic behavior of all three compounds and provide conclusive evidence for bulk superconductivity in CaPd$_2$As$_2$ and SrPd$_2$As$_2$ but only filamentary superconductivity in BaPd$_2$As$_2$ which has a different crystal structure.  The superconducting transition temperatures as estimated from the zero-field $C_{\rm p}(T)$ data are $T_{\rm c} = 1.27(3)$~K for CaPd$_2$As$_2$ and $T_{\rm c} = 0.92(5)$~K for SrPd$_2$As$_2$. The heat capacity jump at $T_{\rm c}$, $\Delta C_{\rm e}(T_{\rm c})$, of CaPd$_2$As$_2$ in $H=0$ is extremely sharp, which allows unambiguous analysis of the derived electronic contribution $C_{\rm e}(T)$ in the superconducting state.  The $\Delta C_{\rm e}(T_{\rm c})/\gamma_{\rm n}T_{\rm c} = 1.14(3)$ is significantly smaller than the BCS prediction of 1.43.  We analyzed the $C_{\rm e}(T)$ data in the superconducting state using the $\alpha$-model,\cite{Padamsee1973,Johnston2013} where $\alpha = \Delta(0)/k_{\rm B}T_{\rm c}$.  A good fit to the data was obtained using $\alpha = 1.58$, which is significantly smaller than the BCS value of~1.764 which we surmise is due to anisotropy in the superconducting $s$-wave gap.  The thermodynamic critical field $H_{\rm c}(T)$ is also in agreement with the $\alpha$-model prediction using the same value of $\alpha$.

While the $\rho (T)$ data of CaPd$_2$As$_2$ exhibit a $T_{\rm c}$ consistent with that obtained from the $C_{\rm p}(T)$ data, the $\rho(T)$ data of SrPd$_2$As$_2$ exhibit a higher $T_{\rm c}$ evidently due to filamentary non-bulk superconductivity.  Our analysis of the normal- and superconducting-state $\rho (T, H)$ and $C_{\rm p}(T, H)$ data of these two compounds and estimated superconducting parameters indicate  type-II superconductivity with small thermodynamic critical fields and upper critical fields that are much smaller than those of the doped FeAs-based superconductors.

\paragraph*{Note Added.} After submission of our paper, an electronic structure study of ${\rm (Ca,Sr,Ba)Pd_2As_2}$ appeared which largely substantiates our analyses of our experimental data for these compounds and offers additional perspectives.\cite{Nekrasov2013}

\acknowledgments

This research was supported by the U.S. Department of Energy, Office of Basic Energy Sciences, Division of Materials Sciences and Engineering.  Ames Laboratory is operated for the U.S. Department of Energy by Iowa State University under Contract No.~DE-AC02-07CH11358.


\appendix*

\section{Presentation and Analysis of $M(H)$ Isotherms}

\subsection{${\rm CaPd_2As_2}$}

\begin{figure}
\includegraphics[width=3.3in]{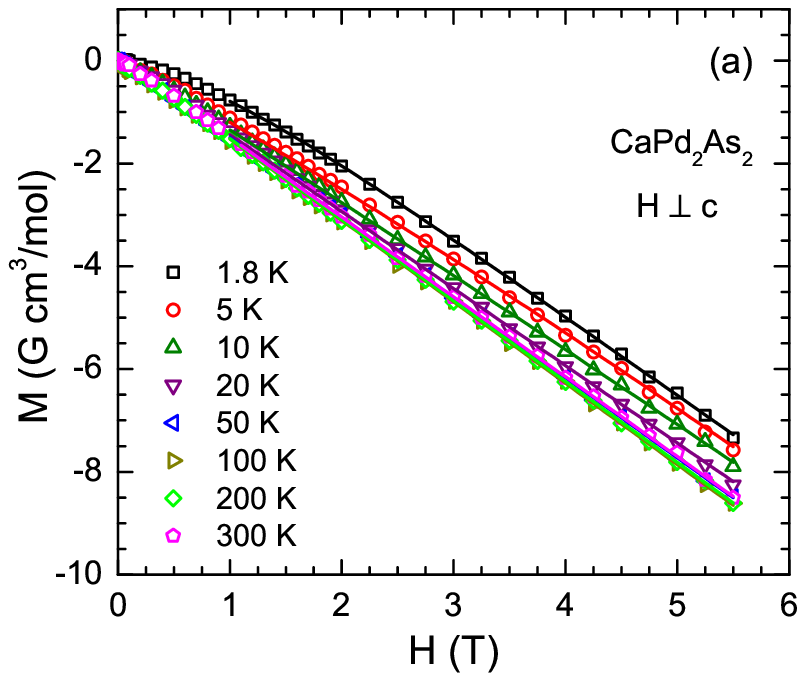}
\includegraphics[width=3.3in]{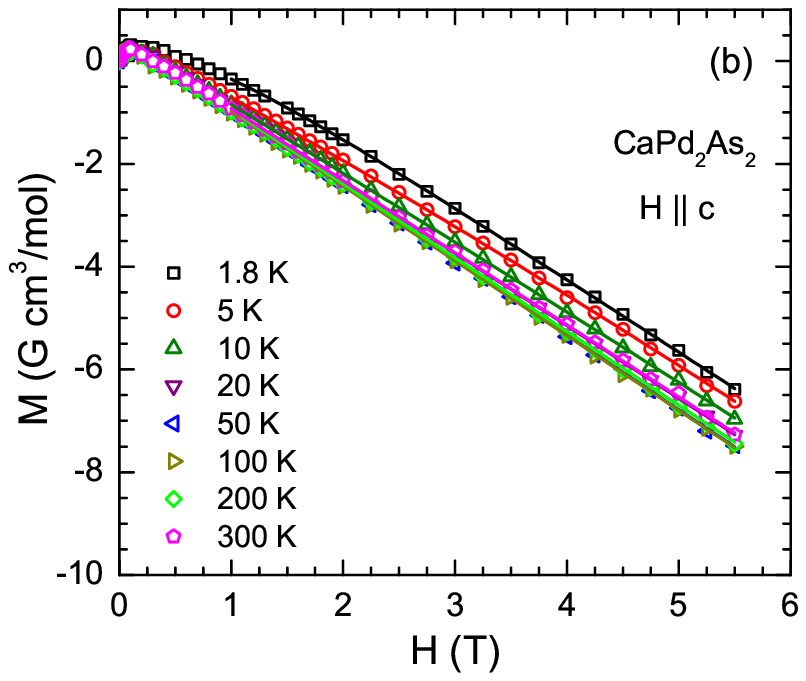}
\caption{(Color online) Isothermal magnetization $M$ of CaPd$_2$As$_2$ versus applied magnetic field $H$ at different temperatures, as listed, for magnetic fields applied (a) in the $ab$-plane ($H \perp  c$) and, (b) along the $c$-axis ($H \parallel c$). The solid curves are fits of the $M(H)$ data by Eqs.~(\ref{Eqs:Brillouin}) for $1.0 \leq H \leq 5.5$~T\@.}
\label{fig:MH_CaPd2As2}
\end{figure}

The $M(H)$ isotherms for a ${\rm CaPd_2As_2}$ crystal measured at eight temperatures between 1.8 and 300~K for $H$ applied both along the $c$-axis ($M_c,\ H \parallel c$) and in the $ab$-plane ($M_{ab},\ H \perp  c$) are shown in Fig.~\ref{fig:MH_CaPd2As2}. Consistent with the $\chi$, the $M$ is diamagnetic (negative) and exhibits weak anisotropy with $M_c(H) > M_{ab}(H)$. For $T \geq 50$~K the $M(H)$ curves are almost linear in $H$, however at low-$T$ a slight nonlinearity is observed that can be attributed to the presence of a small amount of saturable paramagnetic (PM) impurities. Further, the presence of trace amount of ferromagnetic (FM) impurities is also inferred from the $M(H)$ curves.

\begin{table}
\caption{\label{tab:tableMH} Parameters obtained from fitting $M(H)$ isotherms of $A$Pd$_2$As$_2$ ($A$ = Ba, Ca, Sr) at 1.8~K by Eqs.~(\ref{eq:MH_linear-fit}) and (\ref{Eqs:Brillouin} ), where $\theta_{\rm imp} \equiv 0$ and and $S_{\rm imp} \equiv 2$.  Here $M_s$ is the saturation magnetization of ferromagnetic impurities, $\chi$ is the intrinsic susceptibility, and $f_{\rm imp}$ is the molar fraction of the paramagnetic impurities.}
\begin{ruledtabular}
\begin{tabular}{lcccc}
Compound & field  & $M_{\rm s}$  &  $\chi$   &	$f_{\rm imp} $ \\
		& direction & (${\rm \frac{G\,cm^3}{mol}}$) & (${\rm 10^{-5}~\frac{cm^3}{mol}}$) & ($10^{-4}$)\\
\hline
CaPd$_2$As$_2$ & $H \perp c$      &  0.06(7)  & $-15.38(5)$ & 1.01(2)  \\				
			   & $H \parallel c$  &  0.48(3)  & $-14.20(3)$ & 0.86(1)   \\	
SrPd$_2$As$_2$ & $H \perp c$      &  0.002(4) & $-2.60(2)$  &   \\				
			   & $H \parallel c$  &  0.02(1)  & $-5.91(2)$  &    \\	
BaPd$_2$As$_2$ & $H \perp c$      &  0.18(3) & $-14.59(4)$  &   \\				
			   & $H \parallel c$  &  0.20(4)  & $-11.07(3)$  &    \\				
			
\end{tabular}
\end{ruledtabular}
\end{table}

We estimated the FM impurity contribution by fitting the $M(H)$ data for $T\geq50$~K and $H \geq 2$~T by
\begin{equation}
M(H) = M_{\rm s} + \chi H,
\label{eq:MH_linear-fit}
\end{equation}
where $M_{\rm s}$ is the FM impurity saturation magnetization. For $T\geq50$~K the $M_{\rm s}$ is found to be almost $T$-independent and anisotropic with values for $H\parallel c$ and $H\perp c$ listed in Table~\ref{tab:tableMH}.  The  $M_{\rm s}$ value of 0.48~${\rm G\,cm^3/mol}$ for $H\parallel c$ is equivalent to the saturation magnetization of 39~molar ppm of Fe metal impurities suggesting that the only trace amounts of FM impurities are present in the crystal.  However due to the small magnitude of the diamagnetic $\chi$ even trace amounts of FM impurities are observable in $\chi$ and $M$ measurements.

Once the FM impurity contributions $M_{\rm s}$ to the magnetizations were determined, we analyzed the low-$T$ $M(H)$ data for both field directions by
\begin{subequations}
\label{Eqs:Brillouin}
\begin{equation}
M(T,H)=M_{\rm s}+\chi H + f_{\rm imp} M_{\rm{{s_{imp}}}} B_{S_{\rm imp}}(x).
\label{eq:MH_fit}
\end{equation}
Here $\chi$ is the intrinsic susceptibility of the compound, $f_{\rm imp}$ is the molar fraction of PM impurities, $M_{\rm {s_{imp}}} = N_{\rm A} g_{\rm imp} \mu_{\rm B} S_{\rm imp}$ is the PM impurity saturation magnetization, $N_{\rm A}$ is Avogadro's number, $\mu_{\rm B}$ is the Bohr magneton, and $g_{\rm imp}$ and $S_{\rm imp}$ are the spectroscopic splitting factor ($g$-factor) and the spin of the impurities, respectively.  Our unconventional definition of the Brillouin function $B_{S_{\rm imp}}$ is\cite{Johnston2011}
\begin{eqnarray}
B_{S_{\rm imp}}(x) & =& \frac{1}{2S_{\rm imp}}\Bigg\{\left( 2S_{\rm imp}+1 \right) \coth \left[(2S_{\rm imp}+1)\frac{x}{2} \right]  \nonumber\\
				   && \hspace{2cm} -\ \coth \left( \frac{x}{2} \right)\Bigg\},
\label{eq:Brillouin}
\end{eqnarray}
where
\begin{equation}
x \equiv \frac{g_{\rm imp} \mu_{\rm{B}} H}{k_{\rm{B}} (T-\theta_{\rm imp})}.
\end{equation}
\end{subequations}
A Weiss temperature $\theta_{\rm imp}$ is included in the argument of $B_{S_{\rm imp}}(x)$ to take into account for interactions between the paramagnetic impurities in an average mean-field way.

While fitting the $M(H)$ data the impurity $g$-factor was set to $g_{\rm imp}$ = 2 and the $M_{\rm s}$ values for $H \perp c$ and $H \parallel c$ were set to the values listed in Table~\ref{tab:tableMH}. The $M(H)$ data for both $H$ directions were fitted for magnetic fields in the range $1.0 \leq H \leq 5.5$~T\@. The $S_{\rm imp}$ values for both $H$ directions were found to be $S_{\rm imp} = 2.0(2)$; therefore, in the final fits we set $S_{\rm imp} = 2$. The $\theta_{\rm imp}$ values for both field directions were found to be close to zero and therefore in the final fits we set $\theta_{\rm imp} \equiv  0$. The solid curves in Fig.~\ref{fig:MH_CaPd2As2} show the final fits of the $M(H)$ data by Eqs.~(\ref{Eqs:Brillouin}).  The parameters obtained from the fits of the $M(H)$ isotherms at $T=1.8$~K are listed in Table~\ref{tab:tableMH}.  The intrinsic $\chi$ values obtained from the fits of the $M(H)$ data at different temperatures are shown by stars in Fig.~\ref{fig:MT_CaPd2As2}. The $T$ dependence of the intrinsic $\chi$ values clearly indicates that the low-$T$ upturns in the measured $\chi(T)\equiv M(T)/H$ data are extrinsic.

\subsection{${\rm SrPd_2As_2}$}

\begin{figure}
\includegraphics[width=3.3in]{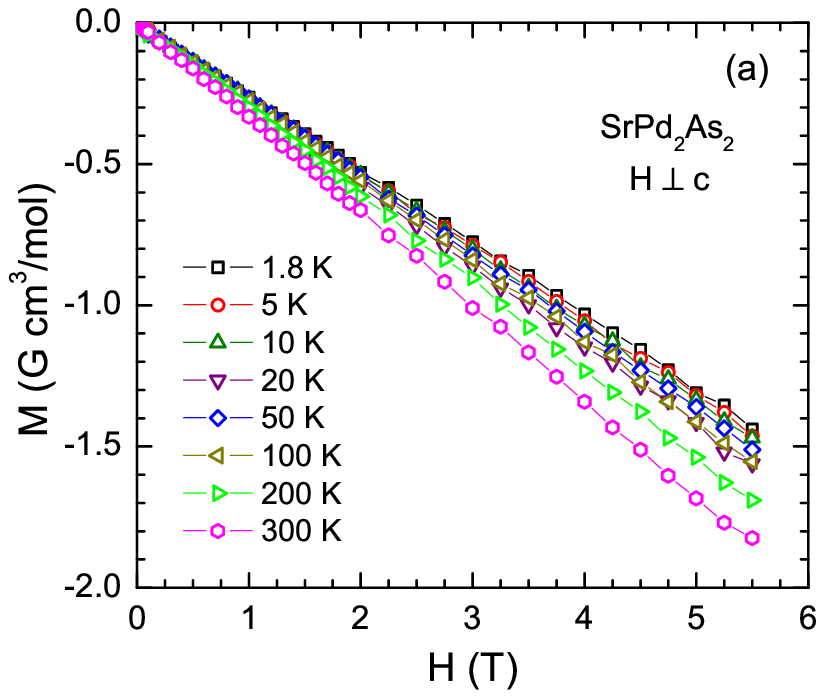}
\includegraphics[width=3.3in]{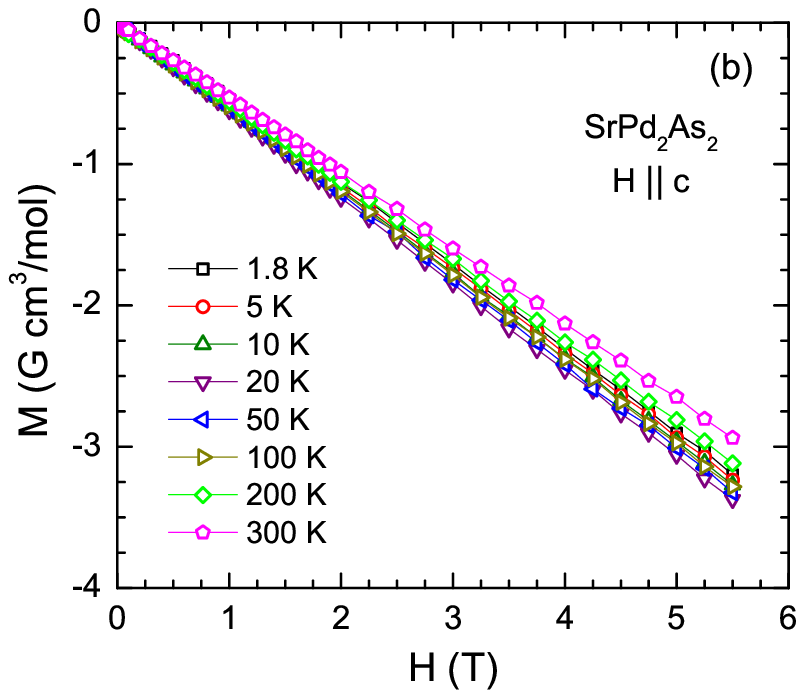}
\caption{\label{fig:MH_SrPd2As2} (Color online) Isothermal magnetization $M$ of SrPd$_2$As$_2$ versus magnetic field $H$ at different temperatures as in Fig.~\ref{fig:MH_CaPd2As2}.}
\end{figure}

The isothermal $M(H)$ data for a SrPd$_2$As$_2$ crystal at different $T$ are shown in Fig.~\ref{fig:MH_SrPd2As2}. Similar to the $\chi(T)\equiv M(T)/H$ data in Fig.~\ref{fig:MT_SrPd2As2}, the $M(H)$ data exhibit anisotropic diamagnetic behavior with $M_{ab}(H) > M_{c}(H)$. In order to obtain the contributions from the FM impurities the $M(H)$ data for $H \geq 2$~T at $T\geq1.8$~K were fitted by Eq.~(\ref{eq:MH_linear-fit}) which gave the average $M_{\rm s}$ values of 0.002(4) for $H \perp  c$ and 0.02(1) for $H \parallel c$. The fitting parameters for both $H \parallel c$ and $H \perp  c$ for $T=1.8$~K are listed in Table~\ref{tab:tableMH}. The intrinsic susceptibilities obtained from the analysis of $M(H)$ data by Eq.~(\ref{eq:MH_linear-fit}) are shown by stars in Fig.~\ref{fig:MT_SrPd2As2}.  The paramagnetic impurity concentration is zero within our resolution, since no curvature in the $M(H)$ curves on cooling to low~$T$ in addition to that due to the FM impurities was detected.

\subsection{${\rm BaPd_2As_2}$}

\begin{figure}
\includegraphics[width=3.3in]{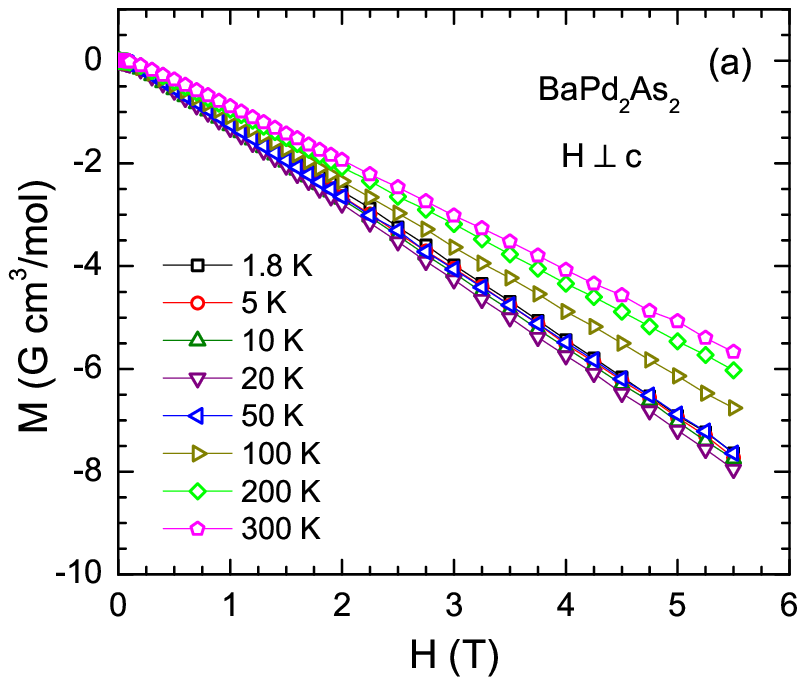}
\includegraphics[width=3.3in]{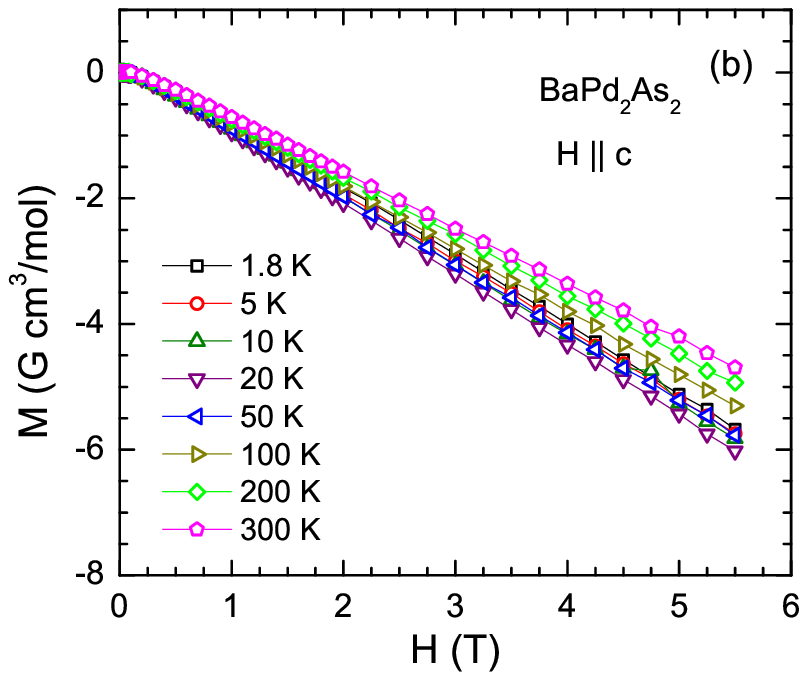}
\caption{\label{fig:MH_BaPd2As2} (Color online) Isothermal magnetization $M(H)$ of BaPd$_2$As$_2$ versus magnetic field measured at different temperatures as in Fig.~\ref{fig:MH_CaPd2As2}.}
\end{figure}

The $M(H)$ isotherms for a ${\rm BaPd_2As_2}$ crtstal at different $T$ are shown in Fig.~\ref{fig:MH_BaPd2As2}. The $M(H)$ curves exhibit weakly anisotropic diamagnetic behavior with $M_{c}(H) > M_{ab}(H)$. The intrinsic $\chi$ was obtained by fitting the $M(H)$ isotherms at each $T$ by Eq.~(\ref{eq:MH_linear-fit}) for $H \geq 2$~T which yielded a temperature-averaged (for $T\geq50$~K) FM saturation values $M_{\rm s}^{ab}$ = 0.18(3) G\,cm$^3$/mol and $M_{\rm s}^{c}$ = 0.20(4) G\,cm$^3$/mol which are equivalent to the magnetization contributions from 12 and 16 molar ppm of Fe metal impurities, respectively.  As in the Sr compound, the paramagnetic impurity concentration is zero within our resolution, since no curvature in the $M(H)$ curves on cooling to low~$T$ in addition to that due to the FM impurities was detected.

\end{document}